\documentclass[12pt]{article}
\usepackage{epsfig}

\begin{document}

\begin{titlepage}

\begin{flushright}
MZ-TH/07-08\\[0.1cm]
June 14, 2007
\end{flushright}

\vspace{1.2cm}
\begin{center}
\Large\bf
Factorization analysis for the fragmentation functions of hadrons containing a heavy quark
\end{center}

\vspace{0.8cm}
\begin{center}
{\sc Matthias Neubert}\footnote{On leave from Laboratory for Elementary-Particle Physics, Cornell University, Ithaca, NY 14853, U.S.A.}\\[0.4cm]
{\sl Institut f\"ur Physik (THEP), Johannes Gutenberg-Universit\"at\\ 
D--55099 Mainz, Germany}
\end{center}

\vspace{1.0cm}
\begin{abstract}
\vspace{0.2cm}
\noindent 
Using methods of effective field theory, a systematic analysis of the fragmentation functions $D_{a/H}(x,m_Q,\mu)$ of a hadron $H$ containing a heavy quark $Q$ is performed (with $a=Q,\bar Q,q,\bar q,g$). By integrating out pair production of virtual and real heavy quarks, the fragmentation functions are matched onto a single nonperturbative function describing the fragmentation of the heavy quark $Q$ into the hadron $H$ in ``partially quenched" QCD. All calculable, short-distance dependence on $x$ is extracted in this step. For $x\to 1$, the remaining fragmentation function can be matched further onto a universal function defined in heavy-quark effective theory in order to factor off its residual dependence on the heavy-quark mass. By solving the evolution equation in the effective theory analytically, large logarithms of the ratio $\mu/m_Q$ are resummed to all orders in perturbation theory. Connections with existing approaches to heavy-quark fragmentation are discussed. In particular, it is shown that previous attempts to extract $\ln^n(1-x)$ terms from the fragmentation function $D_{Q/H}(x,m_Q,\mu)$ are incompatible with a proper separation of short- and long-distance effects.
\end{abstract}
\vfil

\end{titlepage}

\section{Introduction}

The production of heavy quarks (bottom or charm) in high-energy collisions and their subsequent fragmentation into heavy hadrons is an important process in particle physics. For instance, a light Standard Model Higgs boson would preferably decay into a pair of $b\bar b$ quarks, which then fragment into $B$ mesons or heavier beauty states. The direct production of heavy quarks would be an important background process, so a precise calculation of the heavy-quark production and fragmentation processes becomes mandatory. Since the heavy-quark mass provides a natural infrared cutoff, the calculation of the cross section can, to a large extent, be performed using perturbative QCD.

Since the center-of-mass energy $\sqrt s$ of modern particle colliders is much larger than the bottom- or charm-quark masses, the total inclusive heavy-quark production cross section is well approximated by neglecting the masses $m_Q$ of the heavy quarks. However, less inclusive processes such as the differential energy distribution of heavy hadrons produced in $e^+ e^-$ or $pp$ collisions are sensitive to logarithms of the ratio $s/m_Q^2$. The presence of these logarithms threatens the convergence of perturbative expansions. Fortunately, it is possible to show that up to power corrections in $m_Q^2/s$ the cross section factorizes into  mass-independent hard partonic cross sections convoluted with fragmentation functions, which describe the probability that a parton fragments into a particular hadron containing the heavy quark. 

Consider, for concreteness, the production of a heavy meson $H$ containing a heavy quark $Q$ via the decay of a vector boson $V=\gamma, Z$ produced in $e^+ e^-$ annihilation, 
\begin{equation}\label{reaction}
   e^+ e^-\to V(q)\to H(p_H)+X \,, 
\end{equation}
where $s=q^2$. Working in the center-of-mass frame, the energy of the heavy hadron can be measured using the scaling variable
\begin{equation}\label{xdef}
   x = \frac{2p_H\cdot q}{q^2} 
   \stackrel{\rm CMS}{=} \frac{2E_H}{\sqrt s} \,.
\end{equation}
We will be interested in a situation where $s\gg m_Q^2$. At leading power in $m_Q^2/s$, but to all orders in perturbation theory, the differential production cross section can be factorized as \cite{Mele:1990cw} (see also \cite{Collins:1998rz})
\begin{equation}\label{fact}
   \frac{d\sigma_H}{dx} = \sum_a
   \frac{d\hat\sigma_a}{dx}(x,\sqrt{s},\mu)\otimes
   D_{a/H}(x,m_Q,\mu) \,,
\end{equation}
where $D_{a/H}(x,m_Q,\mu)$ gives the probability for a parton $a$ to fragment into a heavy meson $H$ carrying a fraction $x$ of the parton's momentum, and $d\hat\sigma_a/dx$ is the cross section for producing a massless parton $a$ with energy fraction $x$ defined in analogy with (\ref{xdef}). The symbol $\otimes$ denotes the convolution
\begin{equation}
   A(x)\otimes B(x) 
   = \int_0^1\!dz_1\int_0^1\!dz_2\,A(z_1)\,B(z_2)\,\delta(z-z_1 z_2)
   = \int_x^1\!\frac{dz}{z}\,A(z)\,B\Big( \frac{x}{z} \Big) \,.
\end{equation}

The factorization formula (\ref{fact}) separates the dependence on the heavy-quark mass $m_Q$ from the dependence on the center-of-mass energy, which is contained in the partonic cross sections. These hard cross sections can be calculated in the massless approximation. At leading power all mass dependence resides in the fragmentation functions, which are universal (i.e., process-independent) nonperturbative quantities. Like the parton distribution functions, fragmentation functions must be extracted from one set of processes and can then be used to calculate the rates for other processes. Large logarithms of the ratio $s/m_Q^2$ can be resummed to all orders in perturbation theory by evolving the fragmentation functions from a low scale $\mu\sim m_Q$ up to a high scale $\mu\sim\sqrt s$ by solving the DGLAP evolution equations
\begin{equation}\label{DGLAP}
   \frac{d}{d\ln\mu^2}\,D_{a/H}(x,m_Q,\mu)
   = \sum_b P_{a\to b}(x,\mu)\otimes D_{b/H}(x,m_Q,\mu) \,,
\end{equation}
where $P_{a\to b}$ are the time-like Altarelli-Parisi splitting functions \cite{Curci:1980uw,Floratos:1981hs,Collins:1981uw,Mitov:2006ic}. Alternatively, the partonic cross section can be evolved from the high scale down to a scale of order $m_Q$. At leading order in perturbation theory (but not beyond), the functions $P_{a\to b}$ are related to the space-like splitting functions $P_{b\leftarrow a}$ governing the evolution of the parton distribution functions by the Gribov-Lipatov reciprocity relation \cite{Gribov:1971zn}.

In contrast to the case of parton distribution functions, the moments of the fragmentation functions cannot be related to local operator matrix elements. Even the normalization of the leading fragmentation function $D_{Q/H}$ is unknown, since a heavy quark $Q$ can fragment into different heavy hadron states $H$. In the literature on heavy-quark fragmentation one frequently encounters the notion of ``perturbative fragmentation functions" $D_{a/Q}$ \cite{Mele:1990cw}, which describe the probabilities for partons $a$ to fragment into an on-shell heavy quark $Q$. These functions are finite in perturbative QCD, since the heavy-quark mass, or more precisely the scale $m_Q(1-x)$, provides an infrared cutoff. They are relevant for the discussion of inclusive heavy-quark production, where one sums over all possible hadron states $H$ containing the heavy quark $Q$. Indeed, quark-hadron duality suggests that
\begin{equation}
   \sum_H D_{a/H}(x,m_Q,\mu) = D_{a/Q}(x,m_Q,\mu) 
   + O\bigg( \frac{\Lambda_{\rm QCD}}{m_Q(1-x)} \bigg) \,,
\end{equation}
where the sum is over all hadron states containing the heavy quark $Q$. However, such a relation can be expected to hold only if $x$ is not too close to 1, such that the scale $m_Q(1-x)\gg\Lambda_{\rm QCD}$ is in the short-distance regime. The one-loop results for the perturbative fragmentation functions were obtained a long time ago \cite{Mele:1990cw}. They read
\begin{eqnarray}\label{oneloop}
   D_{Q/Q}(x,m_Q,\mu) 
   &=& \delta(1-x) + \frac{C_F\alpha_s}{2\pi}
    \left[ \frac{1+x^2}{1-x} \left( \ln\frac{\mu^2}{m_Q^2(1-x)^2}
    - 1 \right) \right]_+ + O(\alpha_s^2) \,, \nonumber\\
   D_{g/Q}(x,m_Q,\mu) 
   &=& \frac{T_F\alpha_s}{2\pi}
    \left[ x^2 + (1-x)^2 \right]
    \ln\frac{\mu^2}{m_Q^2} + O(\alpha_s^2) \,, 
\end{eqnarray}
while $D_{a/Q}(x,m_Q,\mu)=O(\alpha_s^2)$ for $a\ne Q,g$. Throughout our paper (as is the case in most of the literature on heavy-quark fragmentation) $m_Q$ denotes the pole mass of the heavy quark. If desired, this parameter can be expressed in terms of a short-distance mass parameter using perturbation theory. The two-loop expressions for $D_{a/Q}(x,m_Q,\mu)$ have been calculated in \cite{Melnikov:2004bm,Mitov:2004du}. 

The strong dynamics binding the heavy quark inside a heavy meson implies that there exist important nonperturbative effects leading to a prominent peak of the fragmentation function $D_{Q/H}$ near the endpoint, $1-x_{\rm peak}=O(\Lambda_{\rm QCD}/m_Q)$. For example, the popular model of Peterson et al.\ \cite{Peterson:1982ak}
\begin{equation}\label{Peterson}
   D_{Q/H}(x) 
   = \frac{Nx(1-x)^2}{\left[ (1-x)^2+\epsilon_Q x \right]^2}
\end{equation}
exhibits a peak at $x_{\rm peak}\approx 1-\sqrt{\epsilon_Q}$, where $\epsilon_Q\sim (\Lambda_{\rm QCD}/m_Q)^2$. It is therefore interesting to study the heavy-quark fragmentation process (\ref{reaction}) in the kinematic region $x\to 1$, where $(1-x)$ is treated as a small parameter. In this region the partonic cross sections $d\hat\sigma_a/dx$ contain large logarithms of $(1-x)$, which need to be resummed to all orders in perturbation theory. The traditional approach to resumming these logarithms proceeds via Mellin moment space \cite{Sterman:1986aj,Catani:1989ne}. It has first been applied to the case of the fragmentation process in \cite{Cacciari:2001cw}.\footnote{See also \cite{Mele:1990cw} for a discussion of the complications arising in the $x\to 1$ region.} 
More transparently, it is possible to formulate a factorization theorem for the partonic cross sections using methods of effective field theory and to perform the resummation of large logarithms directly in $x$ space, applying a novel approach developed recently in \cite{Becher:2006nr,Becher:2006mr}. This will be discussed in a forthcoming paper \cite{inprep}. In the $x\to 1$ limit, the fragmentation functions are sensitive to contributions associated with three different mass scales: $m_Q$, $m_Q(1-x)$, and $\Lambda_{\rm QCD}$. The dominant contributions to the cross section come from the region where $m_Q(1-x)\sim\Lambda_{\rm QCD}$, so that it is natural to assign the power counting $(1-x)\sim\Lambda_{\rm QCD}/m_Q$. The goal of the present work is to systematically factorize short- and long-distance effects for the fragmentation functions $D_{a/H}(x,m_Q,\mu)$. The leading singular terms for $x\to 1$ reside in the fragmentation function $D_{Q/H}(x,m_Q,\mu)$ describing the conversion of a heavy quark $Q$ into a heavy hadron $H$ containing that quark.

The problem of how to systematically incorporate and describe  nonperturbative bound-state effects into the heavy-quark fragmentation function has been addressed previously by several authors. The most popular approach is to factorize the fragmentation function into perturbative and nonperturbative components: $D_{Q/H}=D_{Q/H}^{\rm pert}\otimes D_{Q/H}^{np}$ \cite{Colangelo:1992kh,Cacciari:1996wr,Nason:1999zj} (for a recent analysis using this scheme, see e.g.\ \cite{Cacciari:2005uk}). The first component is identified with the perturbative fragmentation function, $D_{Q/H}^{\rm pert}(x,m_Q,\mu)=D_{Q/Q}(x,m_Q,\mu)$, while for the nonperturbative component a model such as (\ref{Peterson}) is adopted. An important observation we make in the present paper is that such an ansatz is {\em incompatible\/} with a proper factorization of short- and long-distance contributions. For $x\to 1$ the perturbative fragmentation function itself contains long-distance contributions from momentum scales of order $m_Q(1-x)\sim\Lambda_{\rm QCD}$, which is evident from the form of the logarithmic term in the first relation in (\ref{oneloop}). Such logarithms are not controllable in perturbation theory and hence must be factorized into the nonperturbative function $D_{Q/H}^{np}$. Previous attempts to resum the $\ln^n(1-x)$ terms in the fragmentation functions have thus led to incorrect formulae for the perturbative Sudakov factors in the $x\to 1$ region. To derive the correct form of the factorization formula for the fragmentation functions is the main objective of our work.

As an aside, we mention that the situation encountered here resembles that in the theoretical analysis of inclusive $B$-meson decays into light hadrons in the ``shape function region". There as well the systematic factorization of short- and long-distance contributions using effective field-theory methods \cite{Bauer:2003pi,Bosch:2004th} has helped to identify some long-distance contributions, which had previously been included in the perturbative decay rate \cite{DeFazio:1999sv}. The proper subtraction of these terms had a significant impact on phenomenology \cite{Lange:2005yw}.

Alternative schemes attempt to model the hadronization process by introducing nonperturbative effects into the (resummed) perturbative fragmentation functions themselves, for instance by using a renormalon analysis \cite{Nason:1996pk,Cacciari:2002xb} or by introducing an effective, infrared-safe low-energy QCD coupling constant \cite{Dokshitzer:1995ev,Aglietti:2006yf}. A different approach was introduced in \cite{Jaffe:1993ie}, where the long-distance effects in the fragmentation process were described using heavy-quark effective theory (HQET). A subsequent work followed up on this idea and aimed at modeling the bound-state effects in the heavy hadron in terms of hadronic form factors \cite{Braaten:1994bz}. The compatibility of the HQET-inspired approaches and those based on perturbative QCD was studied in \cite{Bodwin:2000fd}. Only much later the first systematic matching between the two approaches was performed, and the evolution equation for the HQET fragmentation function was derived \cite{Gardi:2005yi}. In the same paper, a simple relation with the evolution equation for the $B$-meson shape function \cite{Neubert:1993ch,Neubert:1993um} was discovered. In fact, there is a close analogy between the factorization analyses for the heavy-quark fragmentation process and that of inclusive $B$-meson decays into light particles, such as $\bar B\to X_s\gamma$ and $\bar B\to X_u\,l\,\bar\nu$ \cite{Bauer:2003pi,Bosch:2004th,Lange:2005yw}. Using the methods developed in these papers, it would be possible to describe power corrections to the fragmentation functions $D_{a/H}$ in the $x\to 1$ limit in terms of subleading fragmentation functions in HQET, which are analogous to the subleading $B$-meson shape functions studied in \cite{Lee:2004ja,Bosch:2004cb,Beneke:2004in}. Such an analysis is, however, beyond the scope of the present work, and we argue that it would be of little phenomenological relevance.

The remainder of this paper is organized as follows: In the following section we perform a first factorization step at the hard scale $m_Q$ by integrating out real and virtual heavy-quark pairs. This results in a ``partially quenched" version of QCD, in which the heavy quark that fragments into the heavy hadron is still present, but no pair production of heavy quarks is allowed. In this first matching step the different fragmentation functions $D_{a/H}$ are matched onto a single, nonperturbative function describing the fragmentation of the valence 
heavy quark $Q$ into the heavy hadron $H$. This matching step can be performed irrespective of the value of $x$. From a practical point of view, it is the most important result of our analysis, since at this stage all calculable dependence on $x$ has been extracted from the fragmentation functions. In Section~\ref{sec:HQET} we then consider the limit $x\to 1$ and perform a second factorization step at the hard scale $m_Q$, in which ``partially quenched" QCD is matched onto HQET. This is appropriate whenever $m_Q(1-x)\ll m_Q$, and it is a natural step if $m_Q(1-x)\sim\Lambda_{\rm QCD}$, as is the case for the region near the peak of the heavy-quark fragmentation function. We rederive the relation between the perturbative fragmentation function and the perturbative $B$-meson shape function noted in \cite{Gardi:2005yi} and use it to extract the two-loop matching coefficient relating the fragmentation functions defined in HQET and ``partially quenched" QCD. We also present the exact analytic solution of the evolution equation obeyed by the fragmentation function, which is analogous to the solution of the evolution equation for the shape function obtained in \cite{Neubert:2004dd}. In Section~\ref{sec:pheno} we illustrate the phenomenological implications of our results with the help of a simple model for the fragmentation function in HQET. We study the scale dependence of the fragmentation function $D_{Q/H}$ and derive a heavy-quark symmetry relation between the fragmentation functions of $B$ and $D$ mesons. Section~\ref{sec:concl} contains a summary and conclusions. Some technical details of our analysis are relegated to three appendices.

\section{Decoupling of real and virtual heavy-quark pairs}
\label{sec:pairs}

\begin{figure}
\begin{center}
\includegraphics[width=0.85\textwidth]{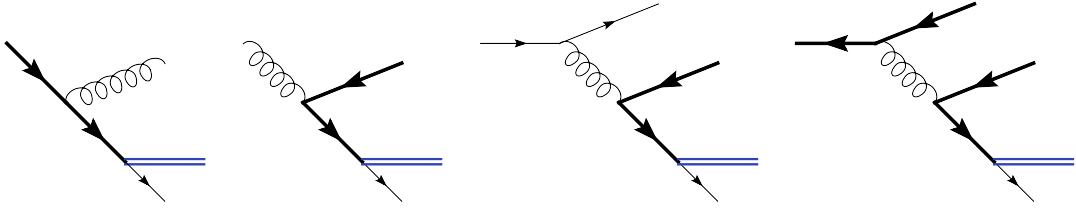}
\end{center}
\vspace{-0.5cm}
\caption{\label{fig:frag}
Examples of fragmentation processes converting a parton $a$ into a heavy hadron $H$. Heavy quarks are denoted by thick lines, while the heavy hadron is drawn as a double line. Only the first process is allowed in ``partially quenched" QCD.}
\end{figure}

The fragmentation functions $D_{a/H}(x,m_Q,\mu)$ receive contributions associated with different mass scales, such as $m_Q$, $m_Q(1-x)$, and $\Lambda_{\rm QCD}$. Our goal is to factorize short-distance effects related to the hard scale $m_Q$ from contributions related to lower scales. It turns out to be useful to integrate out the heavy-quark scale $m_Q$ in two steps. In the first step, QCD with $n_f=n_l+1$ active flavors, in which the factorization formula (\ref{fact}) has been derived, is matched onto a theory with $n_l$ active flavors, which no longer contains $Q\bar Q$ pair production or heavy-quark loops. As always, light flavors are treated as massless. In the $\overline{\rm MS}$ scheme, the matching relation between the running coupling constants $\alpha_s(\mu)$ in the two theories reads
\begin{equation}\label{asmatch}
   \alpha_s^{(n_l+1)}(\mu) = \alpha_s^{(n_l)}(\mu)
   \left[ 1 + \frac23\,T_F\,\frac{\alpha_s^{(n_l)}(\mu)}{2\pi}\,
   \ln\frac{\mu^2}{m_Q^2} + \dots \right] . 
\end{equation}
We would like to write an analogous matching relation for the fragmentation functions $D_{a/H}$. In the theory where the heavy quark $Q$ is among the active flavors, the fragmentation process can be initiated by any parton $a=Q,\bar Q,q,\bar q,g$, as illustrated in Figure~\ref{fig:frag}. However, in all cases but $a=Q$ this will involve the production of a $Q\bar Q$ pair, which is forbidden in the $n_l$-flavor theory. The most general matching relation for the fragmentation functions defined in the two theories therefore reads
\begin{equation}\label{step1}
   D_{a/H}^{(n_l+1)}(x,m_Q,\mu)
   = C_{a/Q}(x,m_Q,\mu)\otimes D_{Q/H}^{(n_l)}(x,m_Q,\mu) \,.
\end{equation}
The meaning of this relation is obvious: the functions $C_{a/Q}$ describe the fragmentation of a massless parton $a$ into a heavy quark $Q$, while $D_{Q/H}^{(n_l)}$ describes the fragmentation of this heavy quark into the heavy hadron $H$. Only the latter process is sensitive to long-distance dynamics.

The functions $C_{a/Q}$ can be extracted from the perturbative expressions for the fragmentation functions derived in \cite{Melnikov:2004bm,Mitov:2004du}. We obtain
\begin{eqnarray}\label{Csing}
   C_{Q/Q}(x,m_Q,\mu) 
   &=& \delta(1-x)
    + c_Q(x) \left( \frac{\alpha_s^{(n_l)}(\mu)}{2\pi} \right)^2 
    + O(\alpha_s^3) \,, \nonumber\\
   C_{g/Q}(x,m_Q,\mu)
   &=& \frac{\alpha_s^{(n_l)}(\mu)}{2\pi}\,T_F
    \left[ x^2 + (1-x)^2 \right]
    \ln\frac{\mu^2}{m_Q^2} 
    + c_g(x) \left( \frac{\alpha_s^{(n_l)}(\mu)}{2\pi} \right)^2 
    + O(\alpha_s^3) \,, \nonumber\\
   C_{a/Q}(x,m_Q,\mu) 
   &=& c_a(x) \left( \frac{\alpha_s^{(n_l)}(\mu)}{2\pi} \right)^2 
    + O(\alpha_s^3) \,; \quad a=\bar Q,q,\bar q.
\end{eqnarray}
To the extent they are known, the expressions for the two-loop coefficients $c_a(x)$ are collected in Appendix~A. These coefficients receive contributions from two sources: terms in the fragmentation functions of the $(n_l+1)$-flavor theory that involve a heavy-quark loop or heavy-quark pair production, and terms resulting from the replacement of the effective coupling constant in the one-loop fragmentation functions according to (\ref{asmatch}). The leading terms in the $x\to 1$ limit reside in $c_Q$ and are given by
\begin{eqnarray}\label{c2}
   c_Q(x) &=& C_F T_F\,\Bigg\{
    \delta(1-x) \left[ \frac12\ln^2\!\frac{\mu^2}{m_Q^2}
    - \left( \frac16 + \frac{2\pi^2}{9} \right) 
    \ln\frac{\mu^2}{m_Q^2} 
    + \frac{3139}{648} - \frac{\pi^2}{3}
    + \frac23\,\zeta_3 \right] \nonumber\\
   &&\mbox{}+ \frac{1}{(1-x)_+} \left( 
    \frac23\ln^2\!\frac{\mu^2}{m_Q^2}
    - \frac{20}{9}\ln\frac{\mu^2}{m_Q^2} 
    + \frac{56}{27} \right) \Bigg\}
    + \mbox{subleading terms.}
\end{eqnarray}
The remaining two-loop matching corrections in (\ref{Csing}) are regular for $x\to 1$. It follows from (\ref{DGLAP}) that the matching coefficients $C_{a/Q}$ and the remaining fragmentation function $D_{Q/H}^{(n_l)}$ obey the evolution equations
\begin{equation}\label{evol2}
   \frac{dC_{a/Q}}{d\ln\mu^2} 
   = \sum_b P_{a\to b}^{(n_l+1)}\otimes C_{b/Q}
    - C_{a/Q}\otimes P_{Q\to Q}^{(n_l)} \,, \qquad
   \frac{dD_{Q/H}^{(n_l)}}{d\ln\mu^2} 
   = P_{Q\to Q}^{(n_l)}\otimes D_{Q/H}^{(n_l)} \,.
\end{equation}
The first relation can be checked using the explicit results for the Wilson coefficients $C_{a/Q}$ given above along with the known expressions for the Altarelli-Parisi splitting functions.

The short-distance coefficients $C_{a/Q}$ can be absorbed into the definition of a partonic heavy-quark production cross section defined in the $n_l$-flavor theory:
\begin{equation}\label{sigdef}
   \frac{d\hat\sigma_Q^{(n_l)}}{dx}(x,\sqrt{s},m_Q,\mu)
   \equiv \sum_a \frac{d\hat\sigma_a^{(n_l+1)}}{dx}(x,\sqrt{s},\mu)
   \otimes C_{a/Q}(x,m_Q,\mu) \,.
\end{equation}
The differential heavy-hadron production cross section can then be written in the simple form
\begin{equation}\label{simplerel}
   \frac{d\sigma_H}{dx} 
   = \frac{d\hat\sigma_Q^{(n_l)}}{dx}(x,\sqrt{s},m_Q,\mu)
   \otimes D_{Q/H}^{(n_l)}(x,m_Q,\mu) \,.
\end{equation}
The advantage is that now only a single fragmentation function remains, which contains all nonperturbative information. The cross section $d\hat\sigma_Q^{(n_l)}/dx$ contains large logarithms, which can be resummed by evolving the partonic cross sections $d\hat\sigma_a^{(n_l+1)}/dx$ from a high scale down to a scale $\mu\sim m_Q$ using the DGLAP evolution equations. 

From a practical point of view, eqs.~(\ref{step1}), (\ref{sigdef}), and (\ref{simplerel}) are the most important results of our work. While in the following section we will factorize the fragmentation function $D_{Q/H}^{(n_l)}(x,m_Q,\mu)$ further by factorizing off its remaining dependence on the heavy-quark mass, all calculable short-distance dependence on $x$ has already been extracted in the stage above. For phenomenological applications it would be appropriate to treat the function $D_{Q/H}^{(n_l)}(x,m_Q,\mu)$ as a hadronic quantity that is extracted from a fit to experimental data. The discussion of the next section becomes relevant only when one tries to relate fragmentation functions describing different processes (see Section~\ref{sec:pheno}).

At this point a comment is in order about a technical detail in the matching calculation which, at first sight, appears problematic. When integrating out heavy-quark pairs we compute short-distance effects associated with the hard scale $m_Q$. However, the resulting Wilson coefficient $c_Q$ in (\ref{c2}) contains a term proportional to $1/(1-x)_+$, which for $x\to 1$ generates a large logarithm. Indeed, an integral of the original fragmentation function over the endpoint region with width $(1-x_0)\sim\Lambda_{\rm QCD}/m_Q$ appears to generate an infrared-sensitive logarithm, 
\begin{eqnarray}
   \int_{x_0}^1\!dx\,D_{Q/H}^{(n_l+1)}(x,m_Q,\mu)
   &\ni& C_F T_F \left( \frac{\alpha_s^{(n_l)}(\mu)}{2\pi} \right)^2 
   \left( \frac23\ln^2\!\frac{\mu^2}{m_Q^2}
    - \frac{20}{9}\ln\frac{\mu^2}{m_Q^2} + \frac{56}{27} \right) 
    \nonumber\\
   &&\times D_{Q/H}^{(n_l)}(x_0,m_Q,\mu) \ln(1-x_0) \,,
\end{eqnarray}
which is not accounted for by the fragmentation function in the low-energy effective theory. But this would be in conflict with the general fact that a low-energy effective theory must be identical to the full theory in the infrared domain. The resolution of the puzzle is that the $\ln(1-x_0)$ term in the relation above should not be interpreted as an infrared-sensitive logarithm arising from the ratio of the two scales $m_Q(1-x)\sim\Lambda_{\rm QCD}$ and $m_Q$, but as an ultraviolet logarithm arising from the presence of the two large scales $s$ and $s(1-x)$ in the fragmentation process (\ref{reaction}), where $s(1-x)$ is the invariant mass squared of the final state $X$, which we assume to be much larger than $m_Q^2$. In fact, these are the {\em only\/} large logarithms that can be extracted from the fragmentation functions in a meaningful way. In a forthcoming paper we will show that the $\mu$-dependent terms in (\ref{c2}) combine with other terms in the resummed expression for the partonic cross section $d\hat\sigma_Q/dx$ in such a way that the running coupling $\alpha_s(\mu)$ is converted from the $(n_l+1)$-flavor to the $n_l$-flavor theory \cite{inprep}. The constant pieces give rise to matching corrections, which are analogous to the matching correction creating a discontinuity in the running coupling constant at three-loop order.

From now on, we focus on the fragmentation function $D_{Q/H}^{(n_l)}$ defined in the theory with $n_l$ active flavors. In the discussion below, quantities without superscript will always be defined in the theory with $n_l$ massless flavors, i.e., we use $\alpha_s\equiv\alpha_s^{(n_l)}$ and $D_{Q/H}\equiv D_{Q/H}^{(n_l)}$.

\section{Factorization for \boldmath $x\to 1$\unboldmath}
\label{sec:HQET}

The step of integrating out virtual and real heavy-quark pairs discussed in the previous section leads to the factorization formula (\ref{step1}), in which the different fragmentation functions of full QCD are related to a single nonperturbative function $D_{Q/H}$ defined in the ``partially quenched" $n_l$-flavor theory. This first matching step can be performed irrespective of the value of $x$. In particular, relation (\ref{simplerel}) for the fragmentation cross section is exact. 

We now return to the specific case of the endpoint region, where $(1-x)\sim\Lambda_{\rm QCD}/m_Q$ is treated as a small parameter. We wish to factorize effects associated with the short-distance scale $m_Q$, which are still present in the $n_l$-flavor theory, from long-distance effects associated with the scales $m_Q(1-x)$ and $\Lambda_{\rm QCD}$. HQET is the appropriate effective theory for describing these nonperturbative  effects \cite{Neubert:1993mb}. Matching ``partially quenched" $n_l$-flavor QCD onto HQET we obtain a matching relation of the form
\begin{equation}\label{match2}
   D_{Q/H}(x,m_Q,\mu)
   \to C_D(m_Q,\mu) \left\{ S_{Q/H}(\hat\omega,\mu)
   + \frac{1}{m_Q} \sum_i C_i^{(1)}(m_Q,\mu)\,
   S_i^{(1)}(\hat\omega,\mu) + \dots \right\} ,
\end{equation}
where $\hat\omega=O(\Lambda_{\rm QCD})$ is a momentum variable in the effective theory yet to be defined. The Wilson coefficients $C_D$ and $C_i^{(n)}$ depend on the heavy-quark mass logarithmically through the running coupling $\alpha_s(m_Q)$. The functions $S_{Q/H}$ and $S_i^{(n)}$ are effective, leading and subleading fragmentation functions defined in HQET (see below). The contributions of the subleading functions are suppressed with respect to the leading term by powers of $\Lambda_{\rm QCD}/m_Q$ or $(1-x)$. For most of this section we will be concerned with the leading term.

We employ the conventional definition of the (bare) fragmentation function (defined in $d=4-2\epsilon$ space-time dimensions) of a heavy quark $Q$ transforming into a heavy hadron $H$ with light-cone momentum fraction $x$, given by \cite{Collins:1981uw}
\begin{equation}\label{Ddef}
   D_{Q/H}(x,m_Q,\mu) 
   = \frac{x^{d-3}}{2\pi} \int dt\,e^{ip\cdot n t}
   \sum_\chi \frac{\rlap{\hspace{0.03cm}/}n_{\alpha\beta}}{2}
   \langle 0|(\psi_Q U_n^*)_\beta^i(tn)|H(p_H)\chi\rangle
   \langle H(p_H)\chi|( U_n^T\bar\psi_Q)_\alpha^i(0)|0\rangle \,.
\end{equation}
Here $n^\mu$ is a light-like vector, $\alpha,\beta$ are Dirac indices, and $i$ is a color index. Longitudinal boost invariance implies that the result should be unchanged under rescalings of the light-like vector $n$. It follows that the right-hand side of (\ref{Ddef}) can only depend on the ratio $x=n\cdot p_H/n\cdot p\in[0,1]$. The expression above is rendered gauge invariant by the insertion of Wilson lines. We define
\begin{equation}\label{Udef}
   U_n(y) = \mbox{P}\exp\left( ig\int_{-\infty}^0\!ds\,
    n\cdot A(y+sn) \right) , \quad
   U_n^\dagger(y) = \overline{\mbox{P}}
    \exp\left( -ig\int_{-\infty}^0\!ds\,
    n\cdot A(y+sn) \right) ,
\end{equation}
where the symbol P means path-ordering, such that gluon fields are ordered from left to right in the order of decreasing $s$ values, and $\overline{\rm P}$ means the opposite ordering. In (\ref{Ddef}) we need the related objects $U_n^*(y)$ and $U_n^T(y)$, which are obtained from $U_n^\dagger(y)$ and $U_n(y)$ by transposition. This reverses the ordering and replaces the gluon fields by the transposed fields $A^T=A_a\,t_a^T$, where $t_a$ are the hermitian generators of color $SU(N_c)$. The transposed Wilson lines appear because in (\ref{Ddef}) the Wilson lines are located on the ``wrong" sides of the quark fields.

The fragmentation function receives contributions from both hard and soft interactions. In order to separate short- and long-distance effects, we match $D_{Q/H}$ onto a corresponding function defined in HQET. Following \cite{Jaffe:1993ie}, we define
\begin{equation}\label{Sdef}
   S_{Q/H}(\hat\omega,\mu) = \frac{n\cdot v}{2\pi} 
   \int dt\,e^{ik\cdot n t} \sum_{\chi_s} 
   \langle 0|(h_v S_n^*)_\alpha^i(tn)|H_\infty(v)\chi_s\rangle
   \langle H_\infty(v)\chi_s|(S_n^T\bar h_v)_\alpha^i(0)|0\rangle \,,
\end{equation}
where $v$ is the four-velocity of the heavy hadron, $h_v(y)$ is the two-component heavy-quark field in HQET \cite{Neubert:1993mb}, and we have used that between these fields $\rlap{\hspace{0.03cm}/}n$ can be replaced with ${n\cdot v}$. The hadron states $H_\infty$ in HQET are taken to have a mass-independent normalization $\langle H_\infty|\bar h_v h_v|H_\infty\rangle=1$ instead of the conventional normalization to $2M_H$ employed in (\ref{Ddef}). In the effective theory we sum over soft states $\chi_s$ (i.e., states having momenta much less than $m_Q$), and the QCD Wilson lines $U_n$, $U_n^\dagger$ are replaced by soft Wilson lines $S_n$, $S_n^\dagger$ defined in analogy with (\ref{Udef}), but with QCD gluon fields replaced with soft gluon fields. Note that $k$ corresponds to the residual momentum of the heavy quark, defined as $p=m_Q v+k$. Boost invariance implies that the right-hand side must be a function of $\omega=n\cdot k/n\cdot v\in[\bar\Lambda,\infty[$, where $\bar\Lambda=M_H-m_Q$ is the residual mass of the heavy hadron state in HQET, and the condition $\omega\ge\bar\Lambda$ follows since the heavy quark must have sufficient energy to produce the heavier hadron state $H$. The above definitions imply the relation $\omega=M_H/x-m_Q$, and indeed one of the main points of \cite{Jaffe:1993ie} was to argue that the fragmentation function should be considered as a function of this particular combination of $x$, $m_Q$ and $M_H$ (see also \cite{Bodwin:2000fd}). We find it more convenient to write the fragmentation function instead as a function of the variable $\hat\omega=\omega-\bar\Lambda$, which takes values between 0 and $\infty$ and obeys the simpler relations
\begin{equation}
   \hat\omega = M_H\,\frac{1-x}{x} \,, \qquad
   x = \frac{M_H}{M_H+\hat\omega} \,.
\end{equation}

After the decoupling of heavy-quark pairs discussed in the previous section, the matching relation between the fragmentation functions in QCD and HQET is local in $x$ and $\hat\omega$, as shown in (\ref{match2}). Using the usual definition of the HQET spinor $h_v$ \cite{Neubert:1993mb} one readily obtains the tree-level relation $D_{Q/H}(x,m_Q,\mu)=M_H\,x\,S_{Q/H}(\hat\omega,\mu)+\dots$, where the factor $M_H$ results from the different normalization of the hadron states in (\ref{Ddef}) and (\ref{Sdef}), and the factor $x$ is due to the $x^{d-3}$ prefactor in (\ref{Ddef}). Beyond tree level, we write the matching relation in the differential form
\begin{equation}\label{match}
   D_{Q/H}(x,m_Q,\mu)\,\frac{dx}{x}
   = g(x)\,C_D(m_Q,\mu)\,S_{Q/H}(\hat\omega,\mu)\,d\hat\omega
   + \mbox{power corrections.}
\end{equation}
Any deviation of the function $g(x)$ from 1 is formally a power correction, so instead of the form $g(x)=x^2$ suggested by the tree-level matching relation given above it would be consistent to set $g(x)=1$ or use any other smooth function obeying the constraint $g(1)=1$. We will see in Section~\ref{sec:residual} that a particularly convenient choice is $g(x)=(1+x^2)/2$.

General arguments analogous to those presented in \cite{Bosch:2004th} suggest that the HQET fragmentation fragmentation vanishes for $\hat\omega\to 0$, which implies that the QCD fragmentation function vanishes for $x\to 1$. Note, however, that in general there is no reason why the fragmentation function should vanish at $x=0$, even though this is built into most phenomenological parameterizations. Indeed, the perturbative results in (\ref{oneloop}) and the corresponding two-loop results in \cite{Melnikov:2004bm,Mitov:2004du} suggest that the fragmentation functions tend to constants modulo logarithms for small~$x$.

\subsection{Matching at one-loop order}
\label{sec:1loop}

\begin{figure}
\begin{center}
\includegraphics[width=0.5\textwidth]{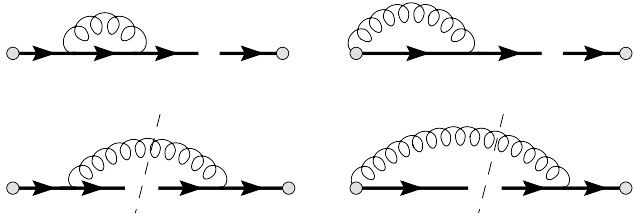}
\end{center}
\vspace{-0.5cm}
\caption{\label{fig:match}
Examples of one-loop diagrams contributing to the perturbative fragmentation functions. The grey dots represent the Wilson lines in expressions (\ref{Ddef}) and (\ref{Sdef}).}
\end{figure}

The hard matching coefficient $C_D$ can be calculated in perturbation theory by computing the fragmentation functions in the two theories and requiring that the two sides in (\ref{match}) be identical up to power-suppressed terms. Since the effective theory (HQET) exactly matches the full theory (``partially quenched" QCD with $n_l$ light flavors) in the infrared, the matching can be performed using on-shell quark and gluon states. Examples of one-loop diagrams contributing to the perturbative fragmentation functions in the two theories are shown in Figure~\ref{fig:match}.

Replacing the heavy hadron by an on-shell heavy quark, we find that at one-loop order in $n_l$-flavor QCD the renormalized fragmentation function $D_{Q/Q}$ in the $\overline{\rm MS}$ scheme is given by the first formula in (\ref{oneloop}). At the parton level we do not distinguish between the heavy-quark mass and the mass of the hadron containing the heavy quark (i.e., we set $M_H=m_Q$). Performing the corresponding on-shell calculation in HQET, we obtain
\begin{equation}\label{SQQ}
   \hat S_{Q/Q}(\hat\omega,\mu) 
   = \delta(\hat\omega) 
   \left( 1 - \frac{C_F\alpha_s}{\pi}\,\frac{\pi^2}{24} \right) 
   - \frac{C_F\alpha_s}{\pi}
   \left[ \frac{1}{\hat\omega} \left( 1 
   + 2\ln\frac{\hat\omega}{\mu} \right) \right]_*^{[\mu]} ,
\end{equation}
where the star distributions are defined as \cite{Bosch:2004th,DeFazio:1999sv}
\begin{equation}\label{stardef}
   \int_0^\Omega\!d\hat\omega\,f(\hat\omega)
   \left( \frac{\ln^n(\hat\omega/\mu)}{\hat\omega} \right)_*^{[\mu]}
   = \int_0^\Omega\!d\hat\omega\,
   \frac{f(\hat\omega)-f(0)}{\hat\omega}\,\ln^n\frac{\hat\omega}{\mu}
    + \frac{f(0)}{n+1}\,\ln^{n+1}\frac{\Omega}{\mu} \,.
\end{equation}
This generalization of the usual plus distributions is required, because in the effective theory there is no reasonable way of defining a dimensionless variable $x\in[0,1]$.
 
The expressions (\ref{oneloop}) and (\ref{SQQ}) must be understood in the sense of distributions. In order to extract the short-distance matching coefficient $C_D$, we integrate these perturbative fragmentation functions with an arbitrary test function over an interval $x\in[m_Q/(m_Q+\omega_0),1]$ and $\hat\omega\in[0,\omega_0]$, respectively, where $\omega_0\ll m_Q$. We then require that the results agree to leading power in $\omega_0/m_Q$. This leads to
\begin{equation}\label{CD1loop}
   C_D(m_Q,\mu) = 1 + \frac{C_F\alpha_s}{2\pi} \left( 
   \frac12\ln^2\!\frac{\mu^2}{m_Q^2} + \frac12\ln\frac{\mu^2}{m_Q^2}
   + 2 + \frac{\pi^2}{12} \right) 
   + O(\alpha_s^2) \,.
\end{equation}
This result agrees with a corresponding expression derived in \cite{Gardi:2005yi}. We stress again that the matching coefficient is free of $\ln(1-x)$ terms. All those terms in the perturbative fragmentation function in (\ref{oneloop}) are of long-distance origin and are still contained in the HQET fragmentation function $S_{Q/H}(\hat\omega,\mu)$.

\subsection{Higher-order matching}
\label{subsec:equality}

It is an intriguing observation that the expression (\ref{SQQ}) for the perturbative HQET fragmentation function can be related to the corresponding perturbative expression for the $B$-meson shape function. The shape function is the HQET analog of the heavy-quark parton distribution function of the hadron $H$ \cite{Neubert:1993ch,Neubert:1993um,Bigi:1993ex}. In analogy with (\ref{Sdef}), it is defined as
\begin{equation}
   S_H(\hat\omega,\mu) 
   = \frac{n\cdot v}{2\pi} \int\!dt\,e^{ik\cdot n t}\,
   \langle H_\infty(v)|(\bar h_v S_n)(0) (S_n^\dagger h_v)(tn)
   |H_\infty(v)\rangle \,.
\end{equation}
In this case $\omega=n\cdot k/n\cdot v$ is restricted to take values between $-\infty$ and $\bar\Lambda$, since the residual light-cone energy of the heavy quark can at most equal the residual mass $\bar\Lambda$ of the heavy hadron.\footnote{Note that in the original variable $\omega=n\cdot k/n\cdot v$ the range of support of the shape function is the complement of the range of support for the HQET fragmentation function.} 
We thus define $\hat\omega=\bar\Lambda-\omega\in[0,\infty[$. When this is done, our one-loop expression in (\ref{SQQ}) coincides with the one-loop expression for the perturbative shape function given in eq.~(33) of \cite{Bosch:2004th}.

The agreement between the perturbative fragmentation and shape functions in HQET, when expressed in terms of the proper variables, is not restricted to one-loop order. Rather, it is an all-order identity between the two functions, which follows from their relation to vacuum expectation values of Wilson loops. This has also been observed in \cite{Gardi:2005yi}. In order to derive this relation we make use of field redefinitions, which decouple the interactions of soft gluons from the effective heavy-quark fields. For the case of the shape function, we use $h_v(y)=S_v(y)\,h_v^{(0)}(y)$ (see, e.g., \cite{Neubert:2005mu}), where the soft Wilson line $S_v$ is defined in analogy with $S_n$, but with the light-like vector $n$ replaced with the time-like vector $v$. For the case of the fragmentation function we need the Wilson lines on the ``wrong" side of the heavy-quark field, so we use instead $h_v(y)=h_v^{(0)}(y)\,S_{-v}^T(y)$. The new fields $h_v^{(0)}$ are ``sterile" and do not couple to any other fields in the theory. The corresponding field operators simply create or annihilate the heavy quarks in the external hadron states, giving rise to free two-component Dirac spinors $u_v$ satisfying $\rlap{\hspace{0.02cm}/}v u_v=u_v$ and $\bar u_v u_v=1$. We define
\begin{equation}
   h_v^{(0)i} |H_\infty(v)\rangle
   = \frac{u_v}{\sqrt N_c}\,|\bar q_H^i(\bar\Lambda v)\rangle \,,
\end{equation}
where the state $|\bar q_H^i(\bar\Lambda v)\rangle$ represents the light degrees of freedom inside the hadron $H$ with color index $i$. This state carries momentum $\bar\Lambda v$, which is what remains from the hadron momentum $M_H v$ when the heavy quark is removed. We may therefore write
\begin{equation}\label{SHexpr}
   S_H(\hat\omega,\mu) 
   = \frac{1}{2\pi N_c} \int\!dt\,
   e^{i(\bar\Lambda-\hat\omega)v\cdot nt}\,
   \langle\bar q_H^i(\bar\Lambda v)|
   \left[ (S_v^\dagger S_n)(0) (S_n^\dagger S_v)(tn) \right]^{ii}
   |\bar q_H^i(\bar\Lambda v)\rangle \,.
\end{equation}
After the decoupling transformation, the corresponding expression for the fragmentation function reads
\begin{equation}\label{SQHexpr}
   S_{Q/H}(\hat\omega,\mu) 
   = \frac{1}{2\pi N_c}
   \int dt\,e^{i(\bar\Lambda+\hat\omega)v\cdot nt} 
   \sum_{\chi_s} \langle 0|(S_{-v}^T S_n^*)^{ij}(tn)
   |\bar q_H^i(\bar\Lambda v)\chi_s\rangle
   \langle\bar q_H^i(\bar\Lambda v)\chi_s|
   (S_n^T S_{-v}^*)^{ji}(0)|0\rangle \,.
\end{equation}
In a partonic picture the spectator quark does not participate in the short-distance process, and to obtain the perturbative shape and fragmentation functions it can simply be dropped from the above matrix elements. At the same time the parameter $\bar\Lambda$ is set to zero, thereby removing the contribution of the spectator quark to the overall momentum balance. This yields for the perturbative shape function
\begin{equation}
   S_Q(\hat\omega,\mu) 
   = \frac{1}{2\pi} \int\!dt\,e^{-i\hat\omega v\cdot nt}\,
    \frac{1}{N_c}\,\langle 0|\,\mbox{Tr}\,(S_v^\dagger S_n)(0)
    (S_n^\dagger S_v)(tn)|0\rangle \,.
\end{equation}
In the case of the perturbative fragmentation function we can now perform the sum over the soft states $\chi_s$, since after the removal of the spectator quark this is a complete sum over states in the effective theory. The result is
\begin{eqnarray}
   S_{Q/Q}(\hat\omega,\mu) 
   &=& \frac{1}{2\pi}
    \int dt\,e^{i\hat\omega v\cdot nt}\,\frac{1}{N_c}\,
    \langle 0|\,\mbox{Tr}\,(S_{-v}^\dagger S_n)^*(tn) 
    (S_n^\dagger S_{-v})^*(0)|0\rangle \nonumber\\
   &=& \frac{1}{2\pi}
    \int dt\,e^{i\hat\omega v\cdot nt}\,\frac{1}{N_c}\,
    \langle 0|\,\mbox{Tr}\,(S_{-v}^\dagger S_n)(tn) 
    (S_n^\dagger S_{-v})(0)|0\rangle^* \nonumber\\
   &=& \frac{1}{2\pi}
    \int dt\,e^{i\hat\omega v\cdot nt}\,\frac{1}{N_c}\,
    \langle 0|\,\mbox{Tr}\,(S_{-v}^\dagger S_n)(0) 
    (S_n^\dagger S_{-v})(tn)|0\rangle \nonumber\\
   &=& \frac{1}{2\pi}
    \int dt\,e^{i\hat\omega v\cdot nt}\,\frac{1}{N_c}\,
    \langle 0|\,\mbox{Tr}\,(S_v^\dagger S_n)(0) 
    (S_n^\dagger S_v)(-tn)|0\rangle 
    = S_Q(\hat\omega,\mu) \,.
\end{eqnarray}
The last step follows since the position-space Wilson loop 
$\langle 0|\,\mbox{Tr}\,(S_{-v}^\dagger S_n)(0)%
(S_n^\dagger S_{-v})(tn)|0\rangle$ is a function of the variable $itn\cdot v$, so replacing the sign of $v$ is equivalent to changing the sign of $t$. It follows that the perturbative fragmentation function coincides with the perturbative shape function. 

Armed with this result, the short-distance coefficient $C_D$ in (\ref{match}) can be computed to two-loop order using existing calculations. The two-loop expression for the fragmentation function  in the ``partially quenched" $n_l$-flavor theory can be extracted from \cite{Melnikov:2004bm} by retaining the leading terms in the $x\to 1$ limit in the expression for the perturbative fragmentation function $D_{Q/Q}(x,m_Q,\mu)$ and accounting for the first matching step (\ref{step1}) using relations (\ref{Csing}) and (\ref{c2}). The integral over the two-loop perturbative fragmentation function in HQET is obtained from the corresponding integral over the two-loop perturbative shape function given in eq.~(38) of \cite{Becher:2005pd}. Matching the two expressions, we find
\begin{eqnarray}\label{CDres}
   C_D(m_Q,\mu) 
   &=& 1 + \frac{C_F\alpha_s(\mu)}{2\pi} \left( 
    \frac{L^2}{2} + \frac{L}{2} + 2 + \frac{\pi^2}{12} \right) \nonumber\\
   &&\mbox{}+ C_F \left( \frac{\alpha_s(\mu)}{2\pi} \right)^2 \Big[
    C_F H_F + C_A H_A + T_F n_f H_f \Big]
    + O(\alpha_s^3) \,,
\end{eqnarray}
where $L=\ln(\mu^2/m_Q^2)$, and the expansion coefficients at two-loop order are
\begin{eqnarray}
   H_F &=& \frac{L^4}{8} + \frac{L^3}{4} 
    + \left( \frac98 + \frac{\pi^2}{24} \right) L^2
    + \left( \frac{11}{8} - \frac{11\pi^2}{24} + 6\zeta_3 \right) L
    \nonumber\\
   &&\mbox{}+ \frac{241}{32} 
    + \left( \frac{13}{12} - 2\ln 2 \right) \pi^2
    - \frac{163\pi^4}{1440} - \frac32\,\zeta_3 \,, \nonumber\\
   H_A &=& \frac{11L^3}{36} 
    + \left( \frac{167}{72} - \frac{\pi^2}{12} \right) L^2
    + \left( \frac{1165}{216} + \frac{7\pi^2}{9} 
    - \frac{15}{2}\,\zeta_3 \right) L
    \nonumber\\
   &&\mbox{}+ \frac{12877}{2592} 
    + \left( \frac{755}{432} + \ln 2 \right) \pi^2
    - \frac{47\pi^4}{720} + \frac{89}{36}\,\zeta_3 \,, \nonumber\\
   H_f &=& - \frac{L^3}{9} - \frac{13L^2}{18}
    - \left( \frac{77}{54} + \frac{2\pi^2}{9} \right) L
    - \frac{1541}{648} - \frac{37\pi^2}{108} 
    - \frac{13}{9}\,\zeta_3 \,.
\end{eqnarray}

If desired, the two matching steps discussed above and in Section~\ref{sec:pairs} can be combined into a single matching relation between the original fragmentation functions $D_{a/H}$ and the HQET fragmentation function $S_{Q/H}$. In the resulting relation, which is of convolution type, the Wilson coefficients $C_{a/Q}$ in (\ref{step1}) are simply multiplied by the coefficient $C_D$ in (\ref{CDres}).

\subsection{Nonperturbative effects}

While the perturbative fragmentation function in HQET coincides with the perturbative shape function to all orders in perturbation theory, such a simple relation will no longer be true nonperturbatively. This can be seen from the different ways in which the spectator quark appears in expressions (\ref{SHexpr}) and (\ref{SQHexpr}). It is also apparent from the fact that the residual momentum $k$ of the heavy quark obeys different kinematical restrictions in the two cases. In the case of the shape function, $n\cdot k$ varies in the range $]-\infty,\bar\Lambda]$ (setting $n\cdot v=1$ for simplicity), and the equation of motion $iv\cdot D\,h_v=0$ of HQET enforces the condition $\langle n\cdot k\rangle=0$ \cite{Neubert:1993ch,Neubert:1993um,Bigi:1993ex}. As a result, the first moment of the shape function is given by $\langle\hat\omega\rangle_{S_H}=\bar\Lambda=M_H-m_Q$, and indeed this relation can be used to define a short-distance, running heavy-quark mass called the ``shape-function mass" to all orders in perturbation theory \cite{Bosch:2004th,Neubert:2004sp}. 

For the fragmentation function, on the other hand, the variable $n\cdot k$ is always positive and varies in the range $[\bar\Lambda,\infty[$, so that the average value $\langle n\cdot k\rangle>\bar\Lambda$ is a nontrivial hadronic parameter. It follows that the first moment of the fragmentation function, $\langle\hat\omega\rangle_{S_{Q/H}}\equiv\epsilon_H>0$, is not related to any known parameter and must be regarded as an unknown hadronic quantity. This, along with the fact that even the normalization of the fragmentation functions is unknown, implies that there is much less theoretical handle on the fragmentation function than there is on the shape function.

Nevertheless, the relation between the perturbative shape and fragmentation functions might be employed to motivate some simple models for the fragmentation function. In the case of the shape function, a renormalon-inspired model based on an analysis of the divergence structure of large-order perturbation theory appears to provide a rather good description of different decay spectra in the inclusive processes $\bar B\to X_s\gamma$ and $\bar B\to X_u\ell\,\bar\nu$ \cite{Gardi:2004ia,Andersen:2005mj}. This model starts from the perturbative shape function and introduces nonperturbative effects through a single model parameter set by the first moment. This suggests that it might be reasonable to use the shape function extracted from the $\bar B\to X_s\gamma$ photon spectrum as a model for the HQET fragmentation function. However, we must account for the fact that the normalizations and the first moments of the two functions are different. As a simple model, we thus propose the form
\begin{equation}
   S_{Q/H}^{\rm model}(\hat\omega,\mu)
   = N_H\,S_H^{\rm model}(\hat\omega,\mu)
   \Big|_{\bar\Lambda\to\epsilon_H} \,.
\end{equation}
A rather flexible, two-parameter model for the shape function, which is known to provide good fits to the experimental data, consists of an exponential times a power of $\hat\omega$ \cite{Kagan:1998ym}. This leads to the form (with $b>0$) 
\begin{equation}\label{ourmodel}
   F(\hat\omega)\equiv S_H^{\rm model}(\hat\omega,\mu)
    \Big|_{\bar\Lambda\to\epsilon_H}
   = \frac{b^b}{\Gamma(b)}\,\frac{\hat\omega^{b-1}}{\epsilon_H^b}\,
   e^{-b\hat\omega/\epsilon_H} \,.
\end{equation}
This function obeys the moment relations
\begin{equation}
   \int_0^\infty\!d\hat\omega\,F(\hat\omega) = 1 \,, \quad
   \int_0^\infty\!d\hat\omega\,\hat\omega\,F(\hat\omega) 
   = \epsilon_H \,, \quad
   \int_0^\infty\!d\hat\omega
    \left( \hat\omega^2 - \langle\hat\omega\rangle^2 \right)
    F(\hat\omega) = \frac{\epsilon_H^2}{b} \,.
\end{equation}
The parameters $\epsilon_H$ and $b$ can be tuned to fit the data for the first and second moment of the fragmentation function. 

\subsection{Residual power corrections}
\label{sec:residual}

So far, we have discussed the leading term in the matching relation (\ref{match}) for the fragmentation function, while in general there is an infinite series of power-suppressed contributions as shown in (\ref{match2}). Like the leading term, these power corrections can be related to nonperturbative, subleading fragmentation functions $S_i^{(n)}$ in HQET. Once a basis of such functions has been constructed, their Wilson coefficients $C_i^{(n)}$ can be determined perturbatively by matching perturbative expressions for the HQET fragmentation functions with the subleading terms in the perturbative fragmentation function $D_{Q/Q}$ (as well as more complicated functions such as $D_{Q/gQ}$) in $n_l$-flavor ``partially quenched" QCD, in analogy with our treatment in Section~\ref{sec:1loop}. 

Using the relations between plus and star distributions collected in Appendix~B, we find that at one-loop order in perturbation theory the power-suppressed terms not accounted for by the leading contribution in the matching relation (\ref{match}) are given by 
\begin{eqnarray}
   && D_{Q/Q}(x,m_Q,\mu) - g(x)\,C_D(m_Q,\mu)\,
    S_{Q/Q}(\hat\omega,\mu)\,x
    \left| \frac{d\hat\omega}{dx} \right| \nonumber\\
   &=& \frac{C_F\alpha_s}{2\pi} \left\{ \frac{2}{1-x}
    \left( \frac{1+x^2}{2} - g(x) \right)
    \left( \ln\frac{\mu^2}{m_Q^2(1-x)^2} - 1 \right) 
    - \frac{4g(x)\ln x}{1-x} \right\} .
\end{eqnarray}
The argument of the logarithm once again indicates that this difference is associated with physics at the low scale $m_Q(1-x)\sim\Lambda_{\rm QCD}$. A particularly simple expression is obtained if we use the special form $g(x)=(1+x^2)/2$ for the weight function on the right-hand side of the matching relation (\ref{match}), since in this case the entire first term vanishes. This particular choice will be adopted in our numerical analysis in Section~\ref{sec:pheno}.

To perform the matching onto subleading fragmentation functions is beyond the scope of the present work. For practical applications of our results, the explicit inclusion of higher-order power corrections is not of much importance due to the fact that even the leading-order function $S_{Q/H}(\hat\omega,\mu)$ is not constrained by any useful conditions. Unlike the case of the shape function, the moments of the fragmentation function are not related to local HQET operators. For practical purposes we may therefore {\em define\/} our model for the function $S_{Q/H}(\hat\omega,\mu)$ to include all power-suppressed terms in (\ref{match2}) -- in any case this model must be tuned to fit experimental data. This definition introduces some subleading dependence on the heavy-quark mass $m_Q$ into the fragmentation function, while the original $S_{Q/H}(\hat\omega,\mu)$ was strictly a universal, $m_Q$-independent function. 

\subsection{Renormalization-group evolution}

As discussed earlier, the fragmentation function $D_{Q/H}$ in the $n_l$-flavor QCD theory obeys the second evolution equation in (\ref{evol2}). The relevant splitting function takes the form \cite{Becher:2006mr}
\begin{equation}\label{PQQsing}
   P_{Q\to Q}(x,\mu) 
   = \frac{\Gamma_{\rm cusp}(\mu)}{(1-x)_+}
   + \gamma^\phi(\mu)\,\delta(1-x) + \mbox{subleading terms,}
\end{equation}
where the regular terms are subleading for $x\to 1$. Here $\Gamma_{\rm cusp}$ is the cusp anomalous dimension of Wilson loops with light-light segments \cite{Korchemskaya:1992je}, which is known to three-loop order \cite{Moch:2004pa}. The quantity $\gamma^\phi$ can be identified with a combination of anomalous dimensions of operators defined in soft-collinear effective theory and is known to the same order \cite{Becher:2006nr,Becher:2006mr,Moch:2004pa}. 

It follows from the discussion of Section~\ref{subsec:equality} that the fragmentation function in HQET obeys the same integro-differential renormalization-group equation as the shape function. It reads
\begin{equation}\label{RGE1}
   \frac{d}{d\ln\mu^2}\,S_{Q/H}(\hat\omega,\mu) 
   = - \int_0^{\hat\omega}\!d\hat\omega'\,
   \Gamma(\hat\omega,\hat\omega',\mu)\,S_{Q/H}(\hat\omega',\mu) \,,
\end{equation}
where to all orders in perturbation theory the kernel has the form \cite{Bosch:2004th,Grozin:1994ni}
\begin{equation}\label{Gam}
   \Gamma(\hat\omega,\hat\omega',\mu)
   = - \Gamma_{\rm cusp}(\alpha_s) 
   \left( \frac{1}{\hat\omega-\hat\omega'} \right)_{\!*}^{[\mu]}
   + \gamma^S(\alpha_s)\,\delta(\hat\omega-\hat\omega') \,.
\end{equation}
The anomalous dimension $\gamma^S$ is known to two-loop order \cite{Gardi:2005yi,Becher:2005pd}. Finally, from (\ref{PQQsing}) and (\ref{Gam}) it follows that the hard matching coefficient obeys the evolution equation
\begin{equation}\label{RGE2}
   \frac{d}{d\ln\mu^2}\,C_D(m_Q,\mu) 
   = \left[ \Gamma_{\rm cusp}(\alpha_s)\,\ln\frac{\mu}{m_Q} 
   + \gamma^\phi(\alpha_s) + \gamma^S(\alpha_s) 
   \right] C_D(m_Q,\mu) \,.
 \end{equation}

The renormalization-group equations (\ref{RGE1}) and (\ref{RGE2}) can be solved analytically and in closed form \cite{Neubert:2004dd}. The solution for the Wilson coefficient reads
\begin{equation}\label{Csol}
   C_D(m_Q,\mu) = \exp\left[ -2S(\mu_h,\mu) 
    - 2a_{\gamma^\phi}(\mu_h,\mu) - 2a_{\gamma^S}(\mu_h,\mu) \right]
   \left( \frac{m_Q}{\mu_h} \right)^{2a_\Gamma(\mu_h,\mu)}
   C_D(m_Q,\mu_h) \,,
\end{equation}
where $\mu_h\sim m_Q$ is a hard matching scale, at which the initial condition $C_D(m_Q,\mu_h)$ is free of large logarithms. The evolution functions are given by\footnote{The Sudakov exponent $S$ should not be confused with the fragmentation function.}
\begin{equation}\label{RGEsols}
   S(\nu,\mu) = - \int\limits_{\alpha_s(\nu)}^{\alpha_s(\mu)}\!
    d\alpha\,\frac{\Gamma_{\rm cusp}(\alpha)}{\beta(\alpha)}
    \int\limits_{\alpha_s(\nu)}^\alpha
    \frac{d\alpha'}{\beta(\alpha')} \,, \qquad
   a_\Gamma(\nu,\mu) = - \int\limits_{\alpha_s(\nu)}^{\alpha_s(\mu)}\!
    d\alpha\,\frac{\Gamma_{\rm cusp}(\alpha)}{\beta(\alpha)} \,, 
\end{equation}
and similarly for the functions $a_{\gamma^\phi}$ and $a_{\gamma^S}$. Here $\beta=d\alpha_s/d\ln\mu$ is the QCD $\beta$-function. The exponential in (\ref{Csol}) accomplishes the resummation of large logarithms to all orders in perturbation theory. The perturbative expansions of the anomalous dimensions are collected in Appendix~C. 

The solution of the integro-differential evolution equation (\ref{RGE1}) has been derived in \cite{Neubert:2004dd} using a technique developed in \cite{Lange:2003ff} (see also \cite{Bosch:2004th,Balzereit:1998yf}). One obtains
\begin{equation}\label{Ssol}
   S_{Q/H}(\hat\omega,\mu) 
   = \exp\left[ 2S(\mu_0,\mu) + 2a_{\gamma^S}(\mu_0,\mu) \right] 
    \frac{e^{-\gamma_E\eta}}{\Gamma(\eta)}
    \int_0^{\hat\omega}\!d\hat\omega'\,
    \frac{S_{Q/H}(\hat\omega',\mu_0)}%
         {\mu_0^\eta(\hat\omega-\hat\omega')^{1-\eta}} \,,
\end{equation}
where $\eta=2a_\Gamma(\mu,\mu_0)$. Here $\mu_0$ serves as a low reference scale still in the perturbative domain, at which a model for the HQET fragmentation function is provided. With the help of the above relation, the fragmentation function can then be evolved to a higher scale $\mu$. It should be stressed that the solutions (\ref{Csol}) and (\ref{Ssol}) are formally independent of the two matching scales $\mu_h$ and $\mu_0$. In practice, a scale dependence remains when one truncates the perturbative expansions of the evolution functions.

The renormalization-group improved expressions for the Wilson coefficient and the HQET fragmentation function can be combined and simplified to obtain the final result for the fragmentation function $D_{Q/H}$ in ``partially quenched" QCD. We find
\begin{eqnarray}\label{Dnice}
   D_{Q/H}(x,m_Q,\mu)
   &=& \frac{M_H\,g(x)}{x}
    \exp\left[ 2S(\mu_0,\mu_h) + 2a_{\gamma^S}(\mu_0,\mu_h) 
    + 2a_{\gamma^\phi}(\mu,\mu_h) \right] 
    \left( \frac{\mu_0}{m_Q} \right)^{2a_\Gamma(\mu,\mu_h)}
    \nonumber\\
   &\times& C_D(m_Q,\mu_h) \,\frac{e^{-\gamma_E\eta}}{\Gamma(\eta)}
    \int_0^{\hat\omega}\!d\hat\omega'\,
    \frac{S_{Q/H}(\hat\omega',\mu_0)}%
         {\mu_0^\eta(\hat\omega-\hat\omega')^{1-\eta}} \,;
    \qquad \eta=2a_\Gamma(\mu,\mu_0) \,,
\end{eqnarray}
where $\hat\omega=M_H(1-x)/x$ on the right-hand side. This relation allows us to derive the fragmentation function $D_{Q/H}$ from a primordial function $S_{Q/H}$ defined at a low renormalization scale.

\section{Phenomenological results}
\label{sec:pheno}

\begin{figure}
\begin{center}
\includegraphics[width=0.9\textwidth]{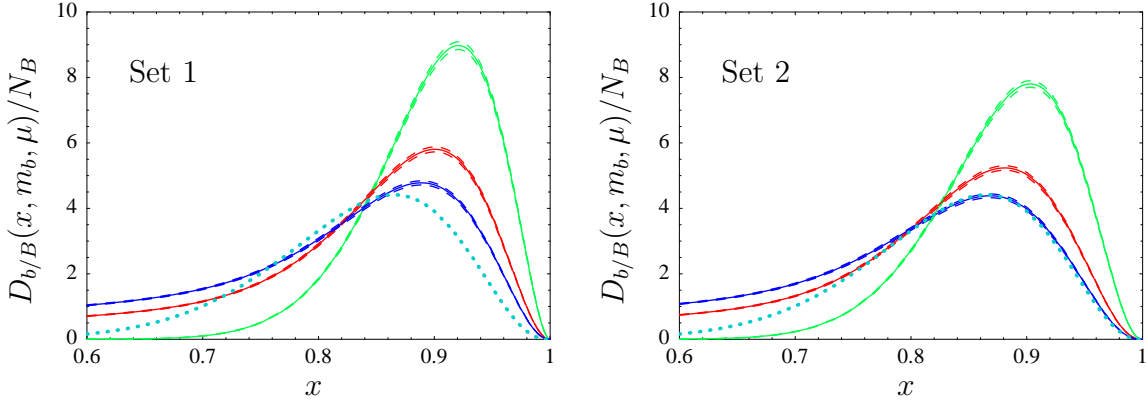}
\end{center}
\vspace{-0.7cm}
\caption{\label{fig:evolution}
Scale dependence of the heavy-quark fragmentation function $D_{b/B}(x,m_b,\mu)$ obtained from (\ref{Dnice}) using the model functions specified in the text. The overall normalization $N_B$ is left free. In each figure, the three sets of curves refer to $\mu=m_b=4.7$\,GeV (blue), $\mu=3$\,GeV (red), and $\mu=\mu_0=1.5$\,GeV (green). Within each set the curves show the residual dependence on the hard matching scale $\mu_h$. The dotted line shows the phenomenological fit at $\mu=m_b$ obtained in \cite{Kniehl:2007yu}.}
\end{figure}

We now illustrate our results for the case of the $B$-meson fragmentation function $D_{b/B}(x,m_b,\mu)$ with the help of a phenomenological model. We use the function (\ref{ourmodel}) with the parameter choices $\epsilon_B=0.69$\,GeV, $b=2.9$ (set~1) and $\epsilon_B=0.82$\,GeV, $b=3.2$ (set~2) to model the HQET fragmentation function $S_{Q/H}(\hat\omega,\mu_0)$ at the scale $\mu_0=1.5$\,GeV. We then use relation (\ref{Dnice}) to obtain the fragmentation function $D_{b/B}(x,m_b,\mu)$ in the $n_l$-flavor theory at a scale $\mu\ge\mu_0$. The evolution from the low scale to a higher scale is performed at next-to-next-to-leading order in renormalization-group improved perturbation theory, using the results collected in Appendix~C. In order to test the stability of the perturbative approximation, we vary the hard matching scale $\mu_h$ between $m_b/\sqrt{2}$ and $\sqrt{2}m_b$. Our results are shown in Figure~\ref{fig:evolution}. We observe that the evolution to higher scales softens the fragmentation function and introduces a significant radiation tail. The dependence on the choice of $\mu_h$ is very weak. Depending on the set of input parameters the functional form of the fragmentation function can be varied. Parameter set~1 provides a good description of the $B$-meson shape function \cite{Bosch:2004th}. Parameter set~2 is chosen such that the resulting fragmentation function at the scale $\mu=m_b$ resembles that obtained from a recent phenomenological fit to $e^+ e^-$ data \cite{Kniehl:2007yu}.

For practical purposes it does not matter whether a model for the fragmentation function is specified at a low or at a high scale. As long as one is interested in, say, $B$-meson fragmentation in $e^+ e^-$ annihilation, one might as well stick with formula (\ref{simplerel}) for the cross section, evolve the partonic cross section down to a scale $\mu=m_b$ (for details see \cite{inprep}), and model the fragmentation function $D_{b/B}(x,\mu=m_b)$ directly. At this stage, all perturbative logarithms of $(1-x)$ have already been extracted and absorbed into the partonic cross section. The factorization of the fragmentation function itself and its scale evolution become important only if one attempts to connect the fragmentation functions extracted in different processes with each other. In particular, treating the charm quark as a heavy quark, it is possible to relate the $b$-quark fragmentation function of $B$ mesons to the charm-quark fragmentation function of $D$ mesons. In comparing the two functions we work for simplicity at next-to-leading order in $\alpha_s$, so that we do not need to worry about the decoupling of real and virtual charm-quark pairs. 

The resulting relation is simplest if we choose to evaluate both fragmentation functions at a common scale $\mu$ and ignore resummation effects. In this case, we obtain at leading power in $\Lambda_{\rm QCD}/m_{c,b}$
\begin{equation}\label{wowsimple}
   D_{b/B}(x,\mu)
   = \frac{C_D(m_b,\mu)}{C_D(m_c,\mu)}\,
   \frac{M_B}{M_D}\,D_{c/D}(y,\mu) \,; \qquad
   y = \frac{M_D}{M_D+M_B\,\frac{1-x}{x}} \,. 
\end{equation}
The relation becomes more complicated when the two functions are evaluated at different scales and resummation effects are taken into account. We then obtain
\begin{equation}\label{wow}
   D_{b/B}(x,\mu=m_b)
   = U(m_c,m_b)\,\frac{M_B}{M_D}\,
   \frac{e^{-\gamma_E\eta}}{\Gamma(\eta)}
   \left( \frac{M_D}{m_c} \right)^\eta
   \int_y^1\frac{dy'}{y^{\prime 2}}\,
   \frac{D_{c/D}(y',\mu=m_c)}%
        {\left(\frac{1-y}{y}-\frac{1-y'}{y'} \right)^{1-\eta}} \,,
\end{equation}
where renormalization-group effects are included in the quantities $\eta=2a_\Gamma(m_b,m_c)$ and
\begin{equation}
   U(m_c,m_b) = \frac{C_D(m_b,m_b)}{C_D(m_c,m_c)}\,
    \exp\left[ 2S(m_c,m_b) + 2a_{\gamma^S}(m_c,m_b) \right] ,
\end{equation}
which we evaluate in 4-flavor QCD. Using $m_b=4.7$\,GeV and $m_c=1.5$\,GeV, the evolution functions evaluate to $\eta\approx 0.322$ and $U(m_c,m_b)\approx 0.965$ at next-to-leading order. 

At present the extraction of the charm-quark fragmentation functions into $D$ mesons appears to be affected by large uncertainties, at least as far as the overall normalization is concerned. For instance, the authors of \cite{Cacciari:2005uk} and \cite{Kniehl:2006mw} find a factor of almost 3 difference in the normalization of the fragmentation functions of $D^+$ and $D^0$ meson. A reason is, perhaps, that a significant fraction of $D$ mesons is produced indirectly via decays of $D^*$ mesons, and these must be subtracted in order to obtain the sample of primary $D$ mesons. From a theoretical perspective, a large difference between the fragmentation functions of $D^+$ and $D^0$ mesons would be very difficult to explain in QCD, as the two functions must coincide up to small electromagnetic and isospin-breaking corrections. We will therefore use the inverse of relation (\ref{wow}) to derive the charm-quark fragmentation function from the fragmentation function of $b$ quarks. The corresponding relation is obtained by replacing $m_c\leftrightarrow m_b$ and $M_D\leftrightarrow M_B$ everywhere. Since the exponent $\eta=2a_\Gamma(m_c,m_b)$ is negative in this case, it is necessary to regularize the integral using a star distribution (see eq.~(47) in \cite{Becher:2006mr} for details). The evolution functions now evaluate to $\eta\approx -0.322$ and $U(m_c,m_b)\approx 0.715$ at next-to-leading order.

Relations (\ref{wowsimple}) and (\ref{wow}) are model-independent consequences of heavy-quark symmetry \cite{Neubert:1993mb}, which are analogous to the relations between semileptonic form factors of $D$ and $B$ mesons obtained by Isgur and Wise \cite{Isgur:1989ed}. Identical relations hold for other pairs of fragmentation functions, such as those for $D^*$ and $B^*$ or $D_s$ and $B_s$ mesons. In practice, we expect these relations to receive sizable power corrections, since the charm quark is not very heavy on the QCD scale. 

\begin{figure}
\begin{center}
\includegraphics[width=7.5cm]{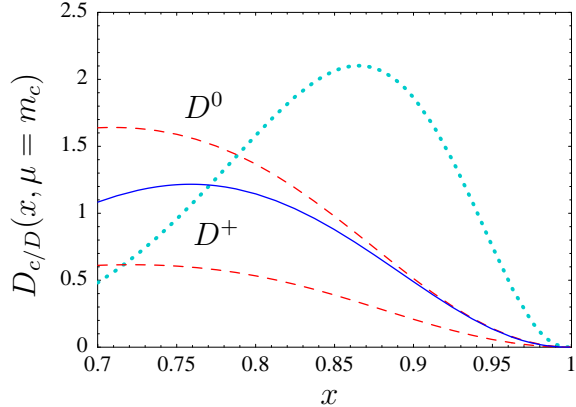}
\end{center}
\vspace{-0.7cm}
\caption{\label{fig:BfromD}
Heavy-quark symmetry prediction for the charm-quark fragmentation function $D_{c/D}(x,\mu=m_c)$ (solid blue line) obtained from the inverse of relation (\ref{wow}) and the phenomenological fit result for the $b$-quark fragmentation function $D_{b/B}(x,\mu=m_b)$ determined in \cite{Kniehl:2007yu} and overlaid as a dotted line. For comparison we show as red dashed lines the fit results for the charm-quark fragmentation functions of $D^+$ and $D^0$ mesons obtained in \cite{Kniehl:2006mw}.}
\end{figure}

To illustrate relation (\ref{wow}) and its inverse we use the result for $D_{b/B}(x,\mu=m_b)$ obtained in \cite{Kniehl:2007yu} from a fit to $e^+ e^-$ data and derive from it a prediction for the charm-quark fragmentation function $D_{c/D}(x,\mu=m_c)$. In Figure~\ref{fig:BfromD}, we compare this prediction with the phenomenological fit results for the charm-quark fragmentation functions of $D^+$ and $D^0$ mesons obtained in \cite{Kniehl:2006mw}. In the region of large $x$ values our prediction comes close to the phenomenological fit for $D^0$ mesons. Note that for values of $x$ below 0.79 the parameter $\hat\omega$ exceeds 0.5\,GeV for the case of charm, so that we would expect significant corrections from subleading fragmentation functions neglected in our analysis.

\section{Conclusions}
\label{sec:concl}

We have performed a systematic factorization analysis for the fragmentation functions of a hadron $H$ containing a heavy quark $Q$. Short- and long-distance contributions associated with different momentum scales have been separated from each other using effective field-theory methods. In several aspects our approach goes beyond previous analyses of heavy-hadron fragmentation functions and introduces some novel features. In particular, we have shown that the popular decomposition of the heavy-quark fragmentation function into perturbative and nonperturbative components, in which the first component is identified with the perturbative fragmentation function $D_{Q/Q}$ in (\ref{oneloop}), does not provide a proper factorization of short- and long-distance effects.

We found it useful to integrate out heavy-quark pair production (virtual and real) by matching the set of fragmentation functions $D_{a/H}(x,m_Q,\mu)$ defined in QCD with $n_l+1$ active flavors onto a {\em single\/} nonperturbative fragmentation function $D_{Q/H}(x,m_Q,\mu)$ defined in ``partially quenched" QCD with $n_l$ active flavors, in which the heavy quark is still present but pair production of heavy quarks is forbidden. Besides the reduction to a single hadronic entity, which offers practical advantages for the modeling of fragmentation functions, in this step all calculable, short-distance dependence on $x$ is extracted from the fragmentation function. We have stressed that most logarithms of $(1-x)$ in the expression for the perturbative fragmentation function $D_{Q/Q}(x,m_Q,\mu)$ are of long-distance origin and thus {\em cannot\/} be extracted and resummed in a meaningful way. We have completed this first matching step at two-loop order with the exception of a single coefficient function, $c_Q(x)$ in (\ref{c2}), for which the terms that are regular for $x\to 1$ have not yet been determined. To complete the calculation of this coefficient is an important task.

In the second part of the paper we have considered the endpoint region, where $(1-x)$ is a small parameter, typically of order $\Lambda_{\rm QCD}/m_Q$. In this case the residual dependence of the fragmentation function on the heavy-quark mass $m_Q$, which remains after the first step, can be extracted by matching ``partially quenched" QCD onto heavy-quark effective theory. We have derived evolution equations for the relevant quantities in the low-energy effective theory and solved them analytically. The resulting relation (\ref{Dnice}) allows us to control the scale dependence of the fragmentation function in the large-$x$ region analytically, and to derive symmetry relations between the fragmentation functions for charm and bottom hadrons. Examples of such relations, which are model-independent consequences of QCD in the heavy-quark limit, have been discussed in Section~\ref{sec:pheno}.

The results obtained in this paper provide the first step in a systematic analysis of fragmentation of heavy hadrons at large $x$ in $e^+ e^-$ or $pp$ collisions. In a forthcoming paper, we will perform the threshold resummation of large logarithms in the fragmentation cross section near $x\to 1$ directly in momentum space. This will set the basis for a precise determination of the functions $D_{b/H}(x,m_b,\mu)$ and $D_{c/H}(x,m_c,\mu)$ from fits to experimental data. Once these functions have been determined at a scale $\mu\sim m_Q$, the entire set of fragmentation functions $D_{a/H}$ can be obtained using relation (\ref{step1}). These functions can then be evolved up to higher scales by solving the DGLAP evolution equations (\ref{DGLAP}).

\subsection*{Acknowledgments}

I am grateful to Alexander Mitov and Hubert Spiesberger for useful discussions, and to Ignazio Scimemi for collaboration during the early stages of this work.

\section*{Appendix~A: Two-loop matching coefficients}

We collect the two-loop expressions for the matching coefficients $C_{a/Q}$ defined in (\ref{step1}) and given in perturbative form in (\ref{Csing}). The two-loop coefficients $c_a$ can be extracted from \cite{Melnikov:2004bm,Mitov:2004du}. Since the analytic expressions obtained in these two papers are lengthy, we will not reproduce them here but rather refer to the equations where they can be found.

We begin with the two-loop coefficient $c_g$ describing the conversion of a massless gluon into a heavy quark $Q$. We obtain
\begin{eqnarray}
   c_g(x) 
   &=& d_g^{(2)}(x) + \frac23\,T_F^2 \left[ x^2 + (1-x)^2 \right]
    \ln^2\!\frac{\mu^2}{m_Q^2} \nonumber\\
   &&\mbox{}- C_F T_F \ln\frac{\mu^2}{m_Q^2}
    \left[ x^2 + (1-x)^2 \right]\otimes \left[ \frac{1+x^2}{1-x}
    \left( \ln\frac{\mu^2}{(1-x)^2 m_Q^2} - 1 \right) \right]_+ ,
\end{eqnarray}
where $d_g^{(2)}(x)$ is given in eq.~(19) of \cite{Mitov:2004du}. The 
second term arises from the application of the matching relation (\ref{asmatch}) for the running coupling constant in the one-loop expression for the gluon fragmentation function, while the convolution term arises from the one-loop cross term in (\ref{step1}). 

The matching coefficients $c_a$ with $a=q,\bar q,\bar Q$ for the conversion of a light quark, light anti-quark, or heavy anti-quark to a heavy quark start at two-loop order and are directly related to the corresponding perturbative fragmentation functions. We have
\begin{equation}
   c_q(x) = c_{\bar q}(x) = d_q^{(2)}(x) \,, \qquad
   c_{\bar Q}(x) = d_{\bar Q}^{(2)}(x) \,,
\end{equation}
where $d_q^{(2)}(x)$ and $d_{\bar Q}^{(2)}(x)$ are given in eqs.~(54) and (56) of \cite{Melnikov:2004bm}, respectively.

\begin{figure}
\begin{center}
\includegraphics[width=0.8\textwidth]{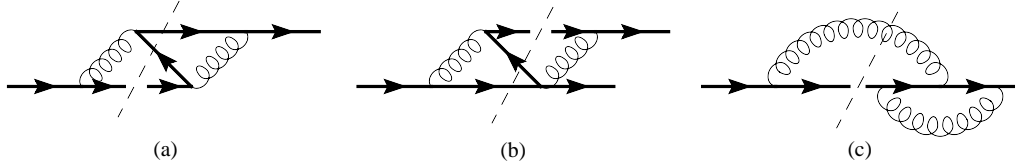}
\end{center}
\vspace{-0.7cm}
\caption{\label{fig:crossed}
(a,b) Diagrams involving heavy-quark pair production, which yield the contribution $\Delta F_Q(x)$ in eq.~(\ref{cQfull}). (c) A diagram with identical color structure, which does not involve pair production and hence does not contribute to $c_Q$.}
\end{figure}

We finally focus on the coefficient $c_Q$, for which we obtain
\begin{eqnarray}\label{cQfull}
   c_Q(x) 
   &=& C_F T_F \left\{ F_Q^{(C_F T_F)}(x) 
    + \frac23\ln\frac{\mu^2}{m_Q^2}
    \left[ \frac{1+x^2}{1-x}
    \left( \ln\frac{\mu^2}{(1-x)^2 m_Q^2} - 1 \right) \right]_+ 
    \right\} \nonumber\\
   &&\mbox{}+ C_F \left( C_F - \frac{C_A}{2} \right)
    \Delta F_Q(x) \,,
\end{eqnarray}
where $F_Q^{(C_F T_F)}(x)$ is given in eq.~(62) of \cite{Melnikov:2004bm}. The term $\Delta F_Q(x)$ contains those parts of the coefficients $F_Q^{(C_F^2)}$ and $F_Q^{(C_A C_F)}$ in eq.~(59) of \cite{Melnikov:2004bm} that are related to the $Q\to QQ\bar Q$ fragmentation channel. They are given by the cuts of the first two diagrams shown in Figure~\ref{fig:crossed}. Due to the presence of other graphs with identical color structure, these contributions cannot be extracted without a dedicated calculation.

\section*{Appendix~B: Relations between distributions}

Using the usual definition of plus distributions together with the definition of the star distributions in (\ref{stardef}), it is possible to derive general relations between the two types of distributions, which are valid for test functions depending on variables $x$ or $\hat\omega$ related by $\hat\omega=m_Q(1-x)/x$. We obtain
\begin{eqnarray}\label{starplus}
   \left( \frac{1}{\hat\omega} \right)_*^{[\mu]} 
   d\hat\omega
   &=& \left\{ \left[ \frac{1}{1-x} \right]_+
    - \ln\frac{\mu}{m_Q}\,\delta(1-x) \right\} \frac{dx}{x} \,, \\
   \left( \frac{\ln(\hat\omega/\mu)}{\hat\omega} \right)_*^{[\mu]} 
   d\hat\omega
   &=& \left\{ \left[ \frac{\ln(1-x)}{1-x} \right]_+ 
    - \ln\frac{\mu}{m_Q} \left[ \frac{1}{1-x} \right]_+ 
    + \frac12\ln^2\!\frac{\mu}{m_Q}\,\delta(1-x) 
    - \frac{\ln x}{1-x} \right\} \frac{dx}{x} \,. \nonumber
\end{eqnarray}

Expressions involving plus distributions of complicated functions of $x$ can be simplified using the identity
\begin{equation}
   \left[ \frac{f(x)}{1-x}\,\ln^n(1-x) \right]_+\!
   = f(x) \left[ \frac{1}{1-x}\,\ln^n(1-x) \right]_+\!
    + \delta(1-x) \int_0^1\!dx\,\frac{f(1)-f(x)}{1-x}\,\ln^n(1-x) \,,
    \quad
\end{equation}
which is valid for integer $n\ge 0$. Applying this result, the one-loop expression for the perturbative heavy-quark fragmentation function in the first equation in (\ref{oneloop}) can be rewritten as
\begin{eqnarray}
   D_{Q/Q}(x,m_Q,\mu) 
   &=& \delta(1-x) \left[ 1 + \frac{C_F\alpha_s}{2\pi} 
    \left( \frac32\ln\frac{\mu^2}{m_Q^2} + 2 \right) \right] 
    \nonumber\\
   &&\mbox{}+ \frac{C_F\alpha_s}{2\pi}\,(1+x^2)
    \left[ \frac{1}{1-x}\!\left( \ln\frac{\mu^2}{m_Q^2(1-x)^2}
     - 1 \right) \right]_+ .
\end{eqnarray}
Using our results (\ref{SQQ}) and (\ref{CD1loop}), and expressing the star distributions in terms of plus distributions with the help of (\ref{starplus}), we find 
\begin{eqnarray}
   g(x)\,C_D(m_Q,\mu)\,S_{Q/Q}(\hat\omega,\mu)\,x
   \left| \frac{d\hat\omega}{dx} \right| 
   &=& \delta(1-x) \left[ 1 + \frac{C_F\alpha_s}{2\pi} 
    \left( \frac32\ln\frac{\mu^2}{m_Q^2} + 2 \right) \right] 
    \\
   &&\mbox{}+ \frac{C_F\alpha_s}{2\pi}\,g(x)
    \left\{ \left[ \frac{2}{1-x}\!\left( \ln\frac{\mu^2}{m_Q^2(1-x)^2}
     - 1 \right) \right]_+\!\! 
     + \frac{4\ln x}{1-x} \right\} . \nonumber
\end{eqnarray}

\section*{Appendix~C: Renormalization-group functions}

The exact solutions (\ref{Csol}) and (\ref{Ssol}) to the evolution equations can be evaluated by expanding the anomalous dimensions and the $\beta$-function as perturbative series in the strong coupling. We work consistently at next-to-next-to-leading order (NNLO) in renormalization-group improved perturbation theory, keeping terms through order $\alpha_s^2$ in the final expressions for the Sudakov exponent $S$ and the functions $a_\Gamma$, $a_{\gamma^\phi}$, and $a_{\gamma^S}$. We define the expansion coefficients as
\begin{eqnarray}
   \Gamma_{\rm cusp}(\alpha_s) &=& \Gamma_0\,\frac{\alpha_s}{4\pi}
    + \Gamma_1 \left( \frac{\alpha_s}{4\pi} \right)^2
    + \Gamma_2 \left( \frac{\alpha_s}{4\pi} \right)^3 + \dots \,,
    \nonumber\\
   \beta(\alpha_s) &=& -2\alpha_s \left[
    \beta_0\,\frac{\alpha_s}{4\pi}
    + \beta_1 \left( \frac{\alpha_s}{4\pi} \right)^2
    + \beta_2 \left( \frac{\alpha_s}{4\pi} \right)^3 + \dots \right] ,
\end{eqnarray}
and similarly for the other anomalous dimensions. In terms of these 
quantities, the NNLO expression for the renormalization-group function $a_\Gamma$ in (\ref{RGEsols}) is given by
\begin{eqnarray}\label{asol}
   a_\Gamma(\nu,\mu)
   &=& \frac{\Gamma_0}{2\beta_0}\,\Bigg\{
    \ln\frac{\alpha_s(\mu)}{\alpha_s(\nu)}
    + \left( \frac{\Gamma_1}{\Gamma_0} - \frac{\beta_1}{\beta_0} 
    \right) \frac{\alpha_s(\mu) - \alpha_s(\nu)}{4\pi} \nonumber\\ 
   &&\mbox{}+ \left[ \frac{\Gamma_2}{\Gamma_0}
    - \frac{\beta_2}{\beta_0} - \frac{\beta_1}{\beta_0}
    \left( \frac{\Gamma_1}{\Gamma_0} - \frac{\beta_1}{\beta_0} 
    \right) \right]
    \frac{\alpha_s^2(\mu) - \alpha_s^2(\nu)}{32\pi^2} + \dots \Bigg\}
    \,.
\end{eqnarray}
The NNLO expression for the Sudakov exponent $S$ in (\ref{RGEsols}) has been given in eq.~(94) of \cite{Becher:2006mr}. We do not reproduce this lengthy formula here.

In the following we list the relevant expansion coefficients for the anomalous dimensions and the QCD $\beta$-function, which are required at NNLO. All results refer to the $\overline{{\rm MS}}$ renormalization scheme. For the convenience of the reader we also quote numerical values corresponding to $n_f=4$ massless quark flavors, as relevant for studies of $b$-quark fragmentation.

The expansion of the cusp anomalous dimension $\Gamma_{\rm cusp}$ to two-loop order was obtained a long time ago \cite{Korchemskaya:1992je}, while the three-loop coefficient has been calculated in \cite{Moch:2004pa}. The results are 
\begin{eqnarray}
   \Gamma_0 &=& 4 C_F = \frac{16}{3} \,, \nonumber\\
   \Gamma_1 &=& 4 C_F \left[ \left( \frac{67}{9} - \frac{\pi^2}{3} 
    \right) C_A - \frac{20}{9}\,T_F n_f \right] 
    \approx 42.7695 \,, \nonumber\\
   \Gamma_2 &=& 4 C_F \Bigg[ C_A^2 \left( \frac{245}{6} 
    - \frac{134\pi^2}{27} + \frac{11\pi^4}{45} 
    + \frac{22}{3}\,\zeta_3 \right) 
    + C_A T_F n_f  \left( - \frac{418}{27} + \frac{40\pi^2}{27}
    - \frac{56}{3}\,\zeta_3 \right) \nonumber\\
   &&\mbox{}+ C_F T_F n_f \left( - \frac{55}{3} + 16\zeta_3 \right) 
    - \frac{16}{27}\,T_F^2 n_f^2 \Bigg] 
    \approx 429.507 \,.
\end{eqnarray}
One also needs the four-loop cusp anomalous dimension, for which the Pad\'e approximation $\Gamma_3\approx 4313$ (for $n_f=4$) was obtained in \cite{Moch:2005ba}.

The anomalous dimension $\gamma^\phi$ is know to three-loop order \cite{Moch:2004pa}. The expansion coefficients are
\begin{eqnarray}
   \gamma_0^\phi &=& 3 C_F = 4 \,, \nonumber\\
   \gamma_1^\phi 
   &=& C_F^2 \left( \frac{3}{2} - 2\pi^2 + 24\zeta_3 \right) 
    + C_F C_A \left( \frac{17}{6} + \frac{22\pi^2}{9} 
    - 12\zeta_3 \right)
    - C_F T_F n_f \left( \frac{2}{3} + \frac{8\pi^2}{9} \right) 
    \approx 43.8275 \,, \nonumber\\
   \gamma_2^\phi
   &=& C_F^3 \left( \frac{29}{2} + 3\pi^2 + \frac{8\pi^4}{5} 
    + 68\zeta_3 - \frac{16\pi^2}{3}\,\zeta_3 - 240\zeta_5 \right)
    \nonumber\\
   &&\mbox{}+ C_F^2 C_A \left( \frac{151}{4} - \frac{205\pi^2}{9}
    - \frac{247\pi^4}{135} + \frac{844}{3}\,\zeta_3
    + \frac{8\pi^2}{3}\,\zeta_3 + 120\zeta_5 \right) \nonumber\\
   &&\mbox{}+ C_F^2 T_F n_f \left( - 46 + \frac{20\pi^2}{9}
    + \frac{116\pi^4}{135} - \frac{272}{3}\,\zeta_3 \right) 
    \nonumber\\
   &&\mbox{}+ C_F C_A^2 \left( - \frac{1657}{36}
    + \frac{2248\pi^2}{81} - \frac{\pi^4}{18} 
    - \frac{1552}{9}\,\zeta_3 + 40\zeta_5 \right) \nonumber\\
   &&\mbox{}+ C_F C_A T_F n_f \left( 40 - \frac{1336\pi^2}{81}
    + \frac{2\pi^4}{45} + \frac{400}{9}\,\zeta_3 \right) \nonumber\\
   &&\mbox{}+ C_F T_F^2 n_f^2 \left( - \frac{68}{9} 
    + \frac{160\pi^2}{81} - \frac{64}{9}\,\zeta_3 \right)
    \approx 617.965 \,.
\end{eqnarray}

The soft anomalous dimension $\gamma^S$ is known to two-loop order \cite{Gardi:2005yi,Becher:2005pd}. Its expansion coefficients are
\begin{eqnarray}
   \gamma_0^S &=& - 2 C_F  = - \frac83 \,, \nonumber\\
   \gamma_1^S &=& C_F \left[ \left( \frac{110}{27}
    + \frac{\pi^2}{18} - 18\zeta_3 \right) C_A
    + \left( \frac{8}{27} + \frac{2\pi^2}{9} \right) T_F n_f \right] 
    \approx -61.4198 \,.
\end{eqnarray}
We use the simple estimate $\gamma_2^S\approx\frac54(\gamma_1^S)^2/\gamma_0^S\approx-1768$ for the three-loop coefficient, as this formula works remarkably well for the known three-loop coefficients of the quantities $\Gamma_{\rm cusp}$, $\gamma^\phi$, and $\beta$.

Finally, the expansion coefficients for the QCD $\beta$-function to four-loop order are
\begin{eqnarray}
   \beta_0 &=& \frac{11}{3}\,C_A - \frac43\,T_F n_f 
    = \frac{25}{3} \,, \nonumber\\
   \beta_1 &=& \frac{34}{3}\,C_A^2 - \frac{20}{3}\,C_A T_F n_f
    - 4 C_F T_F n_f \approx 51.3333 \,, \nonumber\\
   \beta_2 &=& \frac{2857}{54}\,C_A^3 + \left( 2 C_F^2
    - \frac{205}{9}\,C_F C_A - \frac{1415}{27}\,C_A^2 \right) T_F n_f
    + \left( \frac{44}{9}\,C_F + \frac{158}{27}\,C_A 
    \right) T_F^2 n_f^2 \nonumber\\
   &\approx& 406.352 \,, \nonumber\\
   \beta_3 &=& \frac{149753}{6} + 3564\zeta_3
    - \left( \frac{1078361}{162} + \frac{6508}{27}\,\zeta_3 
    \right) n_f
    + \left( \frac{50065}{162} + \frac{6472}{81}\,\zeta_3 
    \right) n_f^2
    + \frac{1093}{729}\,n_f^3 \nonumber\\
    &\approx& 8035.19 \,.
\end{eqnarray}
The value of $\beta_3$ is taken from \cite{vanRitbergen:1997va} and corresponds to $N_c=3$ and $T_F=\frac12$.


\begin{thebibliography}{99}

\bibitem{Mele:1990cw}
  B.~Mele and P.~Nason,
  Nucl.\ Phys.\  B {\bf 361}, 626 (1991).
  
\bibitem{Collins:1998rz}
  J.~C.~Collins,
  Phys.\ Rev.\  D {\bf 58}, 094002 (1998)
  [hep-ph/9806259].

\bibitem{Curci:1980uw}
  G.~Curci, W.~Furmanski and R.~Petronzio,
  Nucl.\ Phys.\  B {\bf 175}, 27 (1980).

\bibitem{Floratos:1981hs}
  E.~G.~Floratos, C.~Kounnas and R.~Lacaze,
  Nucl.\ Phys.\  B {\bf 192}, 417 (1981).

\bibitem{Collins:1981uw}
  J.~C.~Collins and D.~E.~Soper,
  Nucl.\ Phys.\  B {\bf 194}, 445 (1982).

\bibitem{Mitov:2006ic}
  A.~Mitov, S.~Moch and A.~Vogt,
  Phys.\ Lett.\  B {\bf 638}, 61 (2006)
  [hep-ph/0604053].

\bibitem{Gribov:1971zn}
  V.~N.~Gribov and L.~N.~Lipatov,
  Phys.\ Lett.\  B {\bf 37}, 78 (1971).

\bibitem{Melnikov:2004bm}
  K.~Melnikov and A.~Mitov,
  Phys.\ Rev.\  D {\bf 70}, 034027 (2004)
  [hep-ph/0404143].

\bibitem{Mitov:2004du}
  A.~Mitov,
  Phys.\ Rev.\  D {\bf 71}, 054021 (2005)
  [hep-ph/0410205].

\bibitem{Peterson:1982ak}
  C.~Peterson, D.~Schlatter, I.~Schmitt and P.~M.~Zerwas,
  Phys.\ Rev.\  D {\bf 27}, 105 (1983).

\bibitem{Sterman:1986aj}
  G.~Sterman,
  Nucl.\ Phys.\  B {\bf 281}, 310 (1987).

\bibitem{Catani:1989ne}
  S.~Catani and L.~Trentadue,
  Nucl.\ Phys.\  B {\bf 327}, 323 (1989).

\bibitem{Cacciari:2001cw}
  M.~Cacciari and S.~Catani,
  Nucl.\ Phys.\  B {\bf 617}, 253 (2001)
  [hep-ph/0107138].

\bibitem{Becher:2006nr}
  T.~Becher and M.~Neubert,
  Phys.\ Rev.\ Lett.\  {\bf 97}, 082001 (2006)
  [hep-ph/0605050].

\bibitem{Becher:2006mr}
  T.~Becher, M.~Neubert and B.~D.~Pecjak,
  JHEP {\bf 0701}, 076 (2007)
  [hep-ph/0607228].

\bibitem{inprep}
  M. Neubert, 
  in preparation.

\bibitem{Colangelo:1992kh}
  G.~Colangelo and P.~Nason,
  Phys.\ Lett.\  B {\bf 285}, 167 (1992).

\bibitem{Cacciari:1996wr}
  M.~Cacciari, M.~Greco, S.~Rolli and A.~Tanzini,
  Phys.\ Rev.\  D {\bf 55}, 2736 (1997)
  [hep-ph/9608213].

\bibitem{Nason:1999zj}
  P.~Nason and C.~Oleari,
  Nucl.\ Phys.\  B {\bf 565}, 245 (2000)
  [hep-ph/9903541].

\bibitem{Cacciari:2005uk}
  M.~Cacciari, P.~Nason and C.~Oleari,
  JHEP {\bf 0604}, 006 (2006)
  [hep-ph/0510032].

\bibitem{Bauer:2003pi}
  C.~W.~Bauer and A.~V.~Manohar,
  Phys.\ Rev.\  D {\bf 70}, 034024 (2004)
  [hep-ph/0312109].

\bibitem{Bosch:2004th}
  S.~W.~Bosch, B.~O.~Lange, M.~Neubert and G.~Paz,
  Nucl.\ Phys.\  B {\bf 699}, 335 (2004)
  [hep-ph/0402094].

\bibitem{DeFazio:1999sv}
  F.~De Fazio and M.~Neubert,
  JHEP {\bf 9906}, 017 (1999)
  [hep-ph/9905351].

\bibitem{Lange:2005yw}
  B.~O.~Lange, M.~Neubert and G.~Paz,
  Phys.\ Rev.\  D {\bf 72}, 073006 (2005)
  [hep-ph/0504071].

\bibitem{Nason:1996pk}
  P.~Nason and B.~R.~Webber,
  Phys.\ Lett.\  B {\bf 395}, 355 (1997)
  [hep-ph/9612353].

\bibitem{Cacciari:2002xb}
  M.~Cacciari and E.~Gardi,
  Nucl.\ Phys.\  B {\bf 664}, 299 (2003)
  [hep-ph/0301047].

\bibitem{Dokshitzer:1995ev}
  Y.~L.~Dokshitzer, V.~A.~Khoze and S.~I.~Troian,
  Phys.\ Rev.\  D {\bf 53}, 89 (1996)
  [hep-ph/9506425].

\bibitem{Aglietti:2006yf}
  U.~Aglietti, G.~Corcella and G.~Ferrera,
  hep-ph/0610035.

\bibitem{Jaffe:1993ie}
  R.~L.~Jaffe and L.~Randall,
  Nucl.\ Phys.\  B {\bf 412}, 79 (1994)
  [hep-ph/9306201].
  
\bibitem{Braaten:1994bz}
  E.~Braaten, K.~m.~Cheung, S.~Fleming and T.~C.~Yuan,
  Phys.\ Rev.\  D {\bf 51}, 4819 (1995)
  [hep-ph/9409316].

\bibitem{Bodwin:2000fd}
  G.~T.~Bodwin and B.~W.~Harris,
  Phys.\ Rev.\  D {\bf 63}, 077503 (2001)
  [hep-ph/0012037].

\bibitem{Gardi:2005yi}
  E.~Gardi,
  JHEP {\bf 0502}, 053 (2005)
  [hep-ph/0501257].

\bibitem{Neubert:1993ch}
  M.~Neubert,
  Phys.\ Rev.\  D {\bf 49}, 3392 (1994)
  [hep-ph/9311325].

\bibitem{Neubert:1993um}
  M.~Neubert,
  Phys.\ Rev.\  D {\bf 49}, 4623 (1994)
  [hep-ph/9312311].
  
\bibitem{Lee:2004ja}
  K.~S.~M.~Lee and I.~W.~Stewart,
  Nucl.\ Phys.\  B {\bf 721}, 325 (2005)
  [hep-ph/0409045].

\bibitem{Bosch:2004cb}
  S.~W.~Bosch, M.~Neubert and G.~Paz,
  JHEP {\bf 0411}, 073 (2004)
  [hep-ph/0409115].

\bibitem{Beneke:2004in}
  M.~Beneke, F.~Campanario, T.~Mannel and B.~D.~Pecjak,
  JHEP {\bf 0506}, 071 (2005)
  [hep-ph/0411395].

\bibitem{Neubert:2004dd}
  M.~Neubert,
  Eur.\ Phys.\ J.\  C {\bf 40}, 165 (2005)
  [hep-ph/0408179].

\bibitem{Neubert:1993mb}
  M.~Neubert,
  Phys.\ Rept.\  {\bf 245}, 259 (1994)
  [hep-ph/9306320].

\bibitem{Bigi:1993ex}
  I.~I.~Y.~Bigi, M.~A.~Shifman, N.~G.~Uraltsev and A.~I.~Vainshtein,
  Int.\ J.\ Mod.\ Phys.\  A {\bf 9}, 2467 (1994)
  [hep-ph/9312359].

\bibitem{Neubert:2005mu}
  M.~Neubert,
  TASI lectures on ``Effective Field Theory and Heavy Quark Physics", 
  hep-ph/0512222.

\bibitem{Becher:2005pd}
  T.~Becher and M.~Neubert,
  Phys.\ Lett.\  B {\bf 633}, 739 (2006)
  [hep-ph/0512208].

\bibitem{Neubert:2004sp}
  M.~Neubert,
  Phys.\ Lett.\  B {\bf 612}, 13 (2005)
  [hep-ph/0412241].

\bibitem{Gardi:2004ia}
  E.~Gardi,
  JHEP {\bf 0404}, 049 (2004)
  [hep-ph/0403249].

\bibitem{Andersen:2005mj}
  J.~R.~Andersen and E.~Gardi,
  JHEP {\bf 0601}, 097 (2006)
  [hep-ph/0509360].

\bibitem{Kagan:1998ym}
  A.~L.~Kagan and M.~Neubert,
  Eur.\ Phys.\ J.\  C {\bf 7}, 5 (1999)
  [hep-ph/9805303].

\bibitem{Korchemskaya:1992je}
  I.~A.~Korchemskaya and G.~P.~Korchemsky,
  Phys.\ Lett.\  B {\bf 287}, 169 (1992).

\bibitem{Moch:2004pa}
  S.~Moch, J.~A.~M.~Vermaseren and A.~Vogt,
  Nucl.\ Phys.\  B {\bf 688}, 101 (2004)
  [hep-ph/0403192].

\bibitem{Grozin:1994ni}
  A.~G.~Grozin and G.~P.~Korchemsky,
  Phys.\ Rev.\  D {\bf 53}, 1378 (1996)
  [hep-ph/9411323].

\bibitem{Lange:2003ff}
  B.~O.~Lange and M.~Neubert,
  Phys.\ Rev.\ Lett.\  {\bf 91}, 102001 (2003)
  [hep-ph/0303082].

\bibitem{Balzereit:1998yf}
  C.~Balzereit, T.~Mannel and W.~Kilian,
  Phys.\ Rev.\  D {\bf 58}, 114029 (1998)
  [hep-ph/9805297].
  
\bibitem{Kniehl:2007yu}
  B.~A.~Kniehl, G.~Kramer, I.~Schienbein and H.~Spiesberger,
  arXiv:0705.4392 [hep-ph].

\bibitem{Isgur:1989ed}
  N.~Isgur and M.~B.~Wise,
  Phys.\ Lett.\  B {\bf 237}, 527 (1990).

\bibitem{Kniehl:2006mw}
  B.~A.~Kniehl and G.~Kramer,
  Phys.\ Rev.\  D {\bf 74}, 037502 (2006)
  [hep-ph/0607306].

\bibitem{Moch:2005ba}
  S.~Moch, J.~A.~M.~Vermaseren and A.~Vogt,
  Nucl.\ Phys.\  B {\bf 726}, 317 (2005)
  [hep-ph/0506288].

\bibitem{vanRitbergen:1997va}
  T.~van Ritbergen, J.~A.~M.~Vermaseren and S.~A.~Larin,
  Phys.\ Lett.\ B {\bf 400}, 379 (1997)
  [hep-ph/9701390].

\end{thebibliography}
\end{document}